\documentclass{emulateapj}
\usepackage{natbib,aas_macros}
\usepackage{multirow,color}

\begin{document}
\shorttitle{Gas Motion Study of Ly$\alpha$ Emitters}
\shortauthors{Hashimoto et al.}
\slugcomment{Ver. Jan. 9, 2013}

\title{%
Gas Motion Study of Ly$\alpha$ Emitters at $z\sim2$ \\ 
Using FUV and Optical Spectral Lines
\altaffilmark{\dag}
\altaffilmark{\ddag}
}
\author{%
Takuya Hashimoto ~\altaffilmark{1},
Masami Ouchi~\altaffilmark{2,3},
Kazuhiro Shimasaku~\altaffilmark{1,4},
Yoshiaki Ono~\altaffilmark{2}, \\
Kimihiko Nakajima~\altaffilmark{1,3},
Michael Rauch~\altaffilmark{5},
Janice Lee~\altaffilmark{5,6,7},
and
Sadanori Okamura~\altaffilmark{8}
}

\email{thashimoto \_at\_ astron.s.u-tokyo.ac.jp}

\altaffiltext{1}{%
Department of Astronomy, Graduate School of Science, 
The University of Tokyo, Tokyo 113-0033, Japan
}

\altaffiltext{2}{%
Institute for Cosmic Ray Research, The University of Tokyo,
5-1-5 Kashiwanoha, Kashiwa, Chiba 277-8582, Japan
}

\altaffiltext{3}{%
Kavli Institute for the Physics and Mathematics of the Universe,
The University of Tokyo, 5-1-5 Kashiwanoha, Kashiwa, Chiba 277-8583,
Japan
}

\altaffiltext{4}{%
Research Center for the Early Universe, Graduate School of Science,
The University of Tokyo, Tokyo 113-0033, Japan
}

\altaffiltext{5}{%
Observatories of the Carnegie Institution of Washington,
813 Santa Barbara Street, Pasadena, CA 91101, USA
}

\altaffiltext{6}{%
Space Telescope Science Institute, Baltimore, MD, USA
}

\altaffiltext{7}{%
Carnegie Fellow
}

\altaffiltext{8}{%
Department of Advanced Sciences, Faculty of Science and Engineering, 
Hosei University, 3-7-2 Kajino-cho, Koganei-shi, Tokyo 184-8584, 
Japan
}

\altaffiltext{\dag}{%
Some of the data presented herein were obtained at the W. M. Keck 
Observatory, which is operated as a scientific partnership 
among the California Institute of Technology, 
the University of California, 
and the National Aeronautics and Space Administration. 
The Observatory was made possible by the generous financial 
support of the W. M. Keck Foundation.}

\altaffiltext{\ddag}{%
Based in part on data collected at the Subaru Telescope,
which is operated by the National Astronomical Observatory of Japan.}

\slugcomment{Accepted for publication in ApJ, 2013 Jan 9}

\begin{abstract}
We present the results of Magellan/MMIRS and Keck/NIRSPEC
spectroscopy for five Ly$\alpha$ emitters (LAEs) at $z \simeq 2.2$
for which high-resolution FUV spectra from Magellan/MagE 
are available.
We detect nebular emission lines including H$\alpha$
on the individual basis  
and low-ionization interstellar (LIS) absorption lines 
in a stacked FUV spectrum,
and measure average offset velocities of 
the Ly$\alpha$ line, $\Delta v_{\rm Ly\alpha}$, 
and LIS absorption lines, $\Delta v_{\rm abs}$, 
with respect to the systemic velocity
defined by the nebular lines.
For a sample of eight $z \sim 2-3$ LAEs without AGN
from our study and the literature, 
we obtain $\Delta v_{\rm Ly\alpha} = 175\pm35$ km s$^{-1}$, 
which is significantly smaller than that of Lyman-break Galaxies
(LBGs), $\Delta v_{\rm Ly\alpha} \simeq 400$ km s$^{-1}$.
The stacked FUV spectrum gives 
$\Delta v_{\rm abs} = -179 \pm 73$ km s$^{-1}$, 
comparable to that of LBGs.
These positive $\Delta v_{\rm Ly\alpha}$ and 
negative $\Delta v_{\rm abs}$ suggest that LAEs also have 
outflows.
In contrast to LBGs, however, 
the LAEs' $\Delta v_{\rm Ly\alpha}$ is as small as 
$|\Delta v_{\rm abs}|$,
suggesting low neutral hydrogen column densities.
Such a low column density with a small number of resonant scattering 
may cause the observed strong Ly$\alpha$ emission of LAEs.
We find an anti-correlation between
Ly$\alpha$ equivalent width (EW) and $\Delta v_{\rm Ly\alpha}$
in a compilation of LAE and LBG samples.
Although its physical origin is not clear,
this anti-correlation result appears to 
challenge the hypothesis that a strong outflow,
by means of a reduced number of resonant scattering,
produces a large EW. 
If LAEs at $z>6$ have similarly small $\Delta v_{\rm Ly\alpha}$ values,
constraints on the reionization history derived from the Ly$\alpha$ 
transmissivity may need to be revised. 
\end{abstract}



\section{INTRODUCTION} \label{sec:introduction}

Ly$\alpha$ emitters (LAEs) are objects with a large rest-frame 
Ly$\alpha$ equivalent width, ${\rm EW}({\rm Ly}\alpha) \gtrsim 20-30$ \AA.
This population is usually selected using a narrow band filter
for Ly$\alpha$ emission
combined with a broad band filter which measures continuum
emission around Ly$\alpha$.
Very recently, nearby LAEs down to $z \sim 0.2$ have also been
studied with GALEX data \citep[e.g.,][]{cowie2011}.
Previous studies have revealed that most (high-$z$) LAEs are
young, low-mass galaxies with small dust extinction,
while some are old, massive, and dusty \citep[]{ono2010a}.
Morphological studies have revealed that LAEs are smaller than
typical high-$z$ star forming galaxies \citep{bond2009,bond2010}.
Thus, LAEs are among the building block candidates
in the $\Lambda$ Cold Dark Matter (CDM) model, 
where smaller and less massive
galaxies merge to be larger and massive ones \citep{rauch2008}.

Gas exchanges between galaxies and the ambient 
intergalactic medium (IGM),
i.e., outflows and inflows, are thought to play important roles
in galaxy evolution.
Outflows are driven by supernovae (SNe), stellar winds 
from massive stars, and AGN activity 
\citep{heckman1990,murray2005,choi2011}.
Galaxies which lost cold gas via outflows may experience 
a reduction or even termination of subsequent star formation.
In contrast, cold gas supplied by inflows may increase
the star formation, especially when they comes in the form of dense,
filamentary gas streams ({\lq}cold accretion{\rq}; e.g. \citealt{dekel2009}).
Gas exchanges will also affect the chemical evolution of galaxies
and the IGM,
and thus have played important roles in 
 establishing the mass-metalliciy relation across cosmic time
 (e.g., \citealt{larson1974}; \citealt{tremonti2004}; \citealt{erb2006a}).

Outflows have been found in nearby starburst galaxies
\citep[]{heckman1990}, nearby ULIRGs \citep[]{martin2005}, 
Lyman-break galaxies (LBGs)
at $z\sim2-3$ \cite[]{pettini2002,shapley2003}, and 
BX/BM galaxies at $z\sim2$ \citep[]{steidel2010}.
These studies have made use of the fact that 
FUV low-ionization interstellar (LIS) absorption lines,
which are generated when continuum photons encounter
the outflowing gas, are blue-shifted with respect to 
the systemic redshift  measured by nebular emission lines 
such as H$\alpha$ originated from {\sc Hii} regions in the galaxy.

Examining the incidence of outflows in LAEs is of interest, 
because LAEs are generally less massive
($\lesssim 10^{9} M_\odot$; \citealt{nilsson2007, finkelstein2008,
ono2010a,ono2010b,nakajima2012a}) 
than, e.g., LBGs 
($10^{10}$--$10^{11} M_\odot$; \citealt{shapley2004,erb2006b}) 
and 
BzK-selected galaxies 
($10^{10}$--$10^{11} M_\odot$; e.g., \citealt{daddi2004,yuma2012}), 
and thus have shallower gravitational potentials. 
Some authors have argued the
importance of outflows from less massive galaxies in chemical
enrichment \citep[e.g.,][]{larson1974}.
However, FUV continua of LAEs are too faint for LIS absorption
lines to be reliably measured with current facilities.

Ly$\alpha$ is also used to probe the gas kinematics.
The Lya line is known to have complex profiles
caused by its resonant nature.
Many theoretical and observational studies have shown that
outflowing gas leads to a redshifted Ly$\alpha$ line 
with respect to the systemic redshift
\citep[e.g.,][]{verhamme2006,steidel2010},
giving a positive Ly$\alpha$ offset velocity,
$\Delta v_{\rm Ly\alpha}$.
In the case of an outflow, back-scattered (i.e., redshifted) 
Ly$\alpha$ photons have more chance of escape 
because they drop out of resonance with the foreground gas.
However, it is very difficult to obtain high-resolution nebular
emission spectra for LAEs because 
they are faint compared to strong sky emission. 

Prior to our study, only four LAEs have both high-resolution Ly$\alpha$
and nebular line spectra: two from \cite{mclinden2011} and two
from \cite{finkelstein2011b}.
McLinden et al.'s objects have 
$\Delta v _{\rm Ly\alpha}$ = $125\pm17.3$ and $342\pm18.3$ km s$^{-1}$, 
respectively, 
and Finkelstein et al.'s objects have
$\Delta v _{\rm Ly\alpha}$ = $288\pm37\ ({\rm photometric}) 
\pm42\ ({\rm systematic})$ 
and $189\pm35\pm18$ km s$^{-1}$, respectively 
(see also Chonis et al., in preparation). 
The fact that all four have $\Delta v _{\rm Ly\alpha}>0$
suggests that 
outflows are common in LAEs, but due to the small sample 
sizes obtained so far, there has not been a statistical discussion 
of the gas motions of LAEs.

Answering the fundamental question as to
why LAEs have strong Ly$\alpha$ emission is crucial 
for understanding the physical nature of LAEs.
Some studies have shown 
that outflows facilitate the escape of Ly$\alpha$ photons from galaxies
\citep[e.g.,][]{kunth1998} as they reduce the number of resonant
scattering.
Indeed, \cite{verhamme2006,verhamme2008} have carried out
Ly$\alpha$ radiative transfer simulations, and claimed that
expanding shell models can account for observed Ly$\alpha$
spectral profiles of LBGs.
However, while certain aspects of the Ly$\alpha$ radiative transfer 
mechanism of high-$z$ galaxies can be understood
in the context of outflows, the reason for the strong Ly$\alpha$ emission
of LAEs is still an open question.

In order to address these questions, 
we are conducting near infrared spectroscopy
for optically confirmed LAEs at $z\simeq2.2$ promising for detecting
nebular emission lines.
High-resolution spectra of nebular emission lines are 
essential for measuring the offset velocities of the Ly$\alpha$ line, 
$\Delta v_{\rm Ly\alpha}$, 
and of LIS absorption lines, $\Delta v _{\rm abs}$, 
with respect to the systemic velocity.
The redshift of $z \simeq 2.2$ is favored 
since we can simultaneously observe
Ly$\alpha$, 
LIS absorption lines (e.g., {[Si\sc{ii}]}), {[O\sc{i}]}, and {[C\sc{ii}]}),
and optical nebular lines 
(e.g., {\sc [Oii]}, H$\beta$, {\sc [Oiii]}, and H$\alpha$)
from the ground.
Furthermore, we can compare the kinematics and Ly$\alpha$
radiative transfer of LAEs with those of brighter and more
massive galaxies at similar redshifts, e.g., LBGs,
obtained by previous studies.

We successfully detect for several LAEs 
nebular emission lines 
on the individual basis and LIS absorption lines 
in a stacked FUV spectrum,
to measure $\Delta v_{\rm Ly\alpha}$ and $\Delta v _{\rm abs}$.
We also derive physical quantities such as
${\rm EW}({\rm Ly}\alpha)$,
Ly$\alpha$ escape fraction ($f^{{\rm Ly}\alpha}_{\rm esc}$),
star formation rate (SFR), stellar mass, and the size of the star-forming region
from the spectral data and photometric data (SED fitting).
Using these quantities, we discuss the gas kinematics of LAEs
and give implications on the physical origin of the strong 
Ly$\alpha$ emission.

This paper is organized as follows.
Our near infrared spectroscopy is described in \S \ref{sec:data}.
After performing SED fitting in \S \ref{sec:sed},
we derive $\Delta v_{\rm Ly\alpha}$, $\Delta v _{\rm abs}$, 
and several observational quantities for our LAEs 
in \S \ref{sec:physical_quantities}.
A discussion in the context of outflow
and Ly$\alpha$ radiative transfer is given \S \ref{sec:discussion},
followed by conclusions in \S \ref{sec:conclusions}.
Throughout this paper, magnitudes are given in the AB system
\citep{oke1983}, and we assume a $\Lambda$CDM cosmology
with $\Omega_{\small m} = 0.3$, $\Omega_{\small \Lambda} = 0.7$
and $H_{\small 0} = 70$ km s$^{-1}$ Mpc$^{-1}$.

\section{Spectroscopic Data} \label{sec:data}
\subsection{Targets of our NIR Spectroscopy} \label{subsec:sample selection}

Our targets for near infrared spectroscopy are selected from
samples of $z\simeq2.2$ LAEs in the COSMOS and the Chandra Deep Field South
(hereafter CDFS) constructed by K. Nakajima et al. (in preparation) 
in the same manner as in \cite{nakajima2012a}.
These LAE samples are based on narrow-band (NB387) imaging
with Subaru/Suprime-Cam,
supplemented by public broadband data.
NB387 is our custom-made filter with a central wavelength and FWHM of 
$3870$ \AA \ and $94$ \AA, respectively (\citealt{nakajima2012a}).
The LAEs have been selected by imposing 
the following color criteria:
\begin{eqnarray}
 u^{*} - NB387 > 0.5 \ \verb|&| \ B - NB387 > 0.2 &&
{\rm (COSMOS)}, \\
 U - NB387 > 0.8 \  \verb|&| \  B - NB387 > 0.2 &&
 {\rm (CDFS).}
\end{eqnarray}
The COSMOS (CDFS) sample contains 619 (1,108) LAEs
with EW$\gtrsim$30\AA\ down to $NB387=26.1$ (26.4).
Among them, two objects in the CDFS sample (CDFS-3865, CDFS-6482)
and three in the COSMOS sample
(COSMOS-13636, COSMOS-30679, COSMOS-43982)
have a high-quality Magellan/MagE spectrum
of Ly$\alpha$ (M. Rauch et al., in preparation).
These five objects are our targets of NIR spectroscopy.
They have typical NB387 excesses among the whole LAE sample
in each field while having relatively bright NB387 magnitudes.

\subsection{Near Infrared Spectroscopy} \label{subsec:nir_spec}

We observed the two CDFS objects on 2010 October 21
with Magellan/MMIRS using the HK grism covering
$1.254$ -- $2.45\mu$m.
The total exposure time was 10800s for each object.
The slit width was $0.''5$
resulting in $R\equiv \lambda/\Delta\lambda \sim 1120$.
A two-point dither pattern (A1,B1,A2,B2,A3,B3,...) was adopted.
The A0V standard star HIP-16904 was also observed.
The sky was clear through our observation run, with seeing sizes
of $0.''5$ -- $0.''9$.

The three COSMOS objects were observed on 2011 February 10 and 11
with Keck-II/NIRSPEC.
COSMOS-30679 was observed with
NIRSPEC-3 ($J$ band; 1.15 -- 1.36$\mu$m),
NIRSPEC-5 ($H$ band; 1.48 -- 1.76$\mu$m),
and NIRSPEC-6 ($K$ band; 2.2 -- 2.43$\mu$m)
filters in the low-resolution mode,
while COSMOS-13636 and COSMOS-43982 were observed
with the $K$ band alone.
Furthermore, we observed CDFS-3865 with the $J$ band
targeting {\sc [Oii]} $\lambda\lambda3726,3729$.
Total exposure times are shown in 
Table \ref{tab:observation}.
The slit width was $0.''76$ for all three objects,
corresponding to $R \sim 1500$ for all grisms.
A two-point dither pattern was adopted.
We simultaneously observed reference stars, 
which we used for blind offsets (see \citealt{finkelstein2011b,yang2011})
because of the faintness of our targets.
The A0V standard star HIP-13917 was also observed.
The sky was clear in our observing nights, with seeing sizes
of $0.''6$ -- $0.''9$.
A summary of the spectroscopic observations is given in
Table \ref{tab:observation}.

\begin{deluxetable*}{cccccccc}
\tablecolumns{8}
\tablewidth{0pt}
\tablecaption{Summary of our observations \label{tab:observation}}
\tablehead{
\colhead{Object}   & \colhead{$\alpha$(J2000)} & \colhead{$\delta$(J2000)} & \colhead{$z$(Ly$\alpha$)}  
& \colhead{$L$(Ly$\alpha$)} & \colhead{Dates} & \colhead{$t_{\rm exp}$}  \\
\colhead{}    & \colhead{}   & \colhead{} & \colhead{} & \colhead{(10$^{42}$ erg s$^{-1}$)}
   & \colhead{} & \colhead{(s)}\\
   \colhead{(1)}    & \colhead{(2)}   & \colhead{(3)} & \colhead{(4)} & \colhead{(5)}
   & \colhead{(6)} & \colhead{(7)} 
}
\startdata
CDFS-3865
& 03:32:32.31  & $-28$:00:52.20 &  $2.17507^{+0.00104}_{-0.00004}$ & $29.8\pm4.9$
& 2010 Oct 21 & 5100 ($J$), 10800 ($HK$)\\
CDFS-6482
& 03:32:49.34 & $-27$:59:52.35 & $2.20610^{+0.00049}_{-0.00002}$ & $15.4\pm8.1$
& 2010 Oct 21 &  10800 ($HK$) 
\\
COSMOS-13636
& 09:59:59.38 & $+02$:08:38.36 &   $ 2.16229\pm0.00008$ & $11.3\pm0.5$
& 2011 Feb 10--11 &  5400 ($K$) 
\\
 COSMOS-30679
& 10:00:29.81& $+02$:18:49.00 & $2.20046\pm0.00008$ & $8.5\pm0.7$
& 2011 Feb 10--11  &  5400 ($J$), 7200 ($H$), 6300 ($K$)
\\
 COSMOS-43982
& 09:59:54.39 & $+02$:26:29.96 & $2.19396\pm0.00008$  & $11.0\pm0.5$
& 2011 Feb 10--11  &  3600 ($K$)
\enddata

\tablenotetext{}{Notes.
(1) Object ID;
(2), (3) Right ascension and declination;
(4) Redshift of Ly$\alpha$ emission;
(5) Ly$\alpha$ luminosity derived from narrow- and broad-band photometry; 
(6) Dates of observations;
(7) Exposure times for the filters shown in parentheses.
}

\end{deluxetable*}

\subsection{Data Reduction} \label{subsec:reduction}

\begin{figure}[]
\centering
\includegraphics[width=8.8cm]{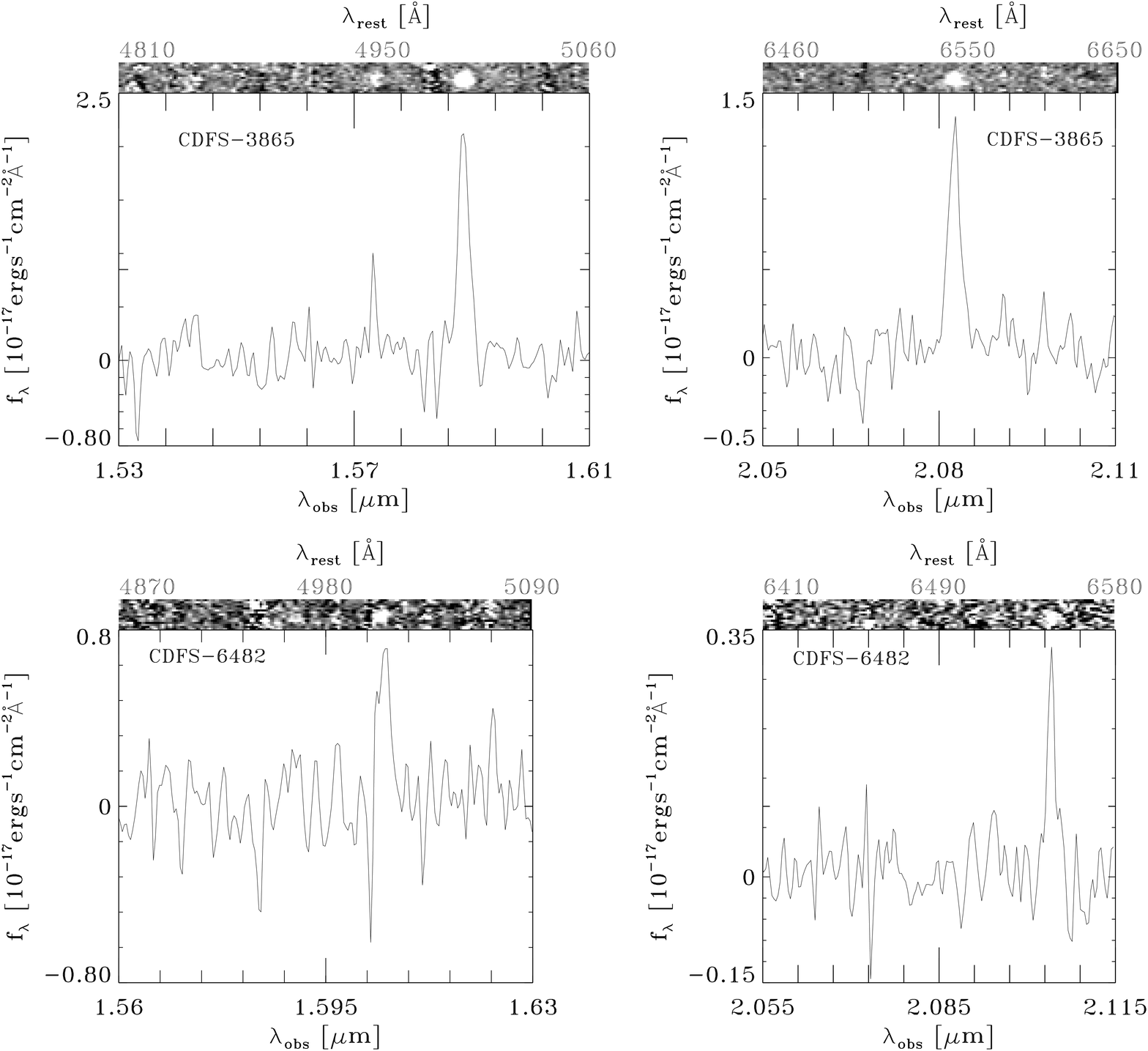}
\caption[]
{
Reduced 1D+2D spectra of CDFS-3865 (top) and CDFS-6482 (bottom).
}
\label{fig:cdfs_spec}
\end{figure}

We reduced the MMIRS data using IRAF tasks and the COSMOS package
which is the standard reduction pipeline for Magellan/IMACS.
The MMIRS detector is read out non-destructively during a single
exposure, and individual read-outs are stored as separate extensions
in a FITS file through which we know whether and when
a particular pixel is saturated.
Bias subtraction and flat fielding were processed for each read-out
using IRAF {\tt mscred} package to treat data of this format.
Then we ran {\tt mmfixen} package which takes advantage of its sampling.
This essentially fits a line to different values for a given pixel
in each readout,
and outputs the slope of this linear fit to the final collapsed image.
Wavelength calibration and distortion correction were processed
for each frame using the COSMOS package.
Although we obtained arc lamp calibration images, we used OH lines
for wavelength calibration.
We then performed the following operation to remove sky background:
 C1 = B1 - (A1 + A2)/2, C2 = B2 - (A2 + A3)/2.
After this operation, we ran {\tt subsky} in the COSMOS package
\footnote{This uses the Kelson procedure \citep[cf,][]{kelson2003}.}
on each frame to remove residual sky lines.
Resultant frames ($C1,C2$,...) are then stacked to have
a final 2D frame using {\tt sumspec-2d} and {\tt extract-2dspec}
in the COSMOS package.
1D spectrum extraction was carried out using {\tt apall} in IRAF.
The telluric absorption correction and flux calibration were
conducted using the standard star frames.
The flux-calibrated 1D spectra of the two objects are shown in
Figure \ref{fig:cdfs_spec}.

\begin{figure}[]
\centering
\includegraphics[width=9cm]{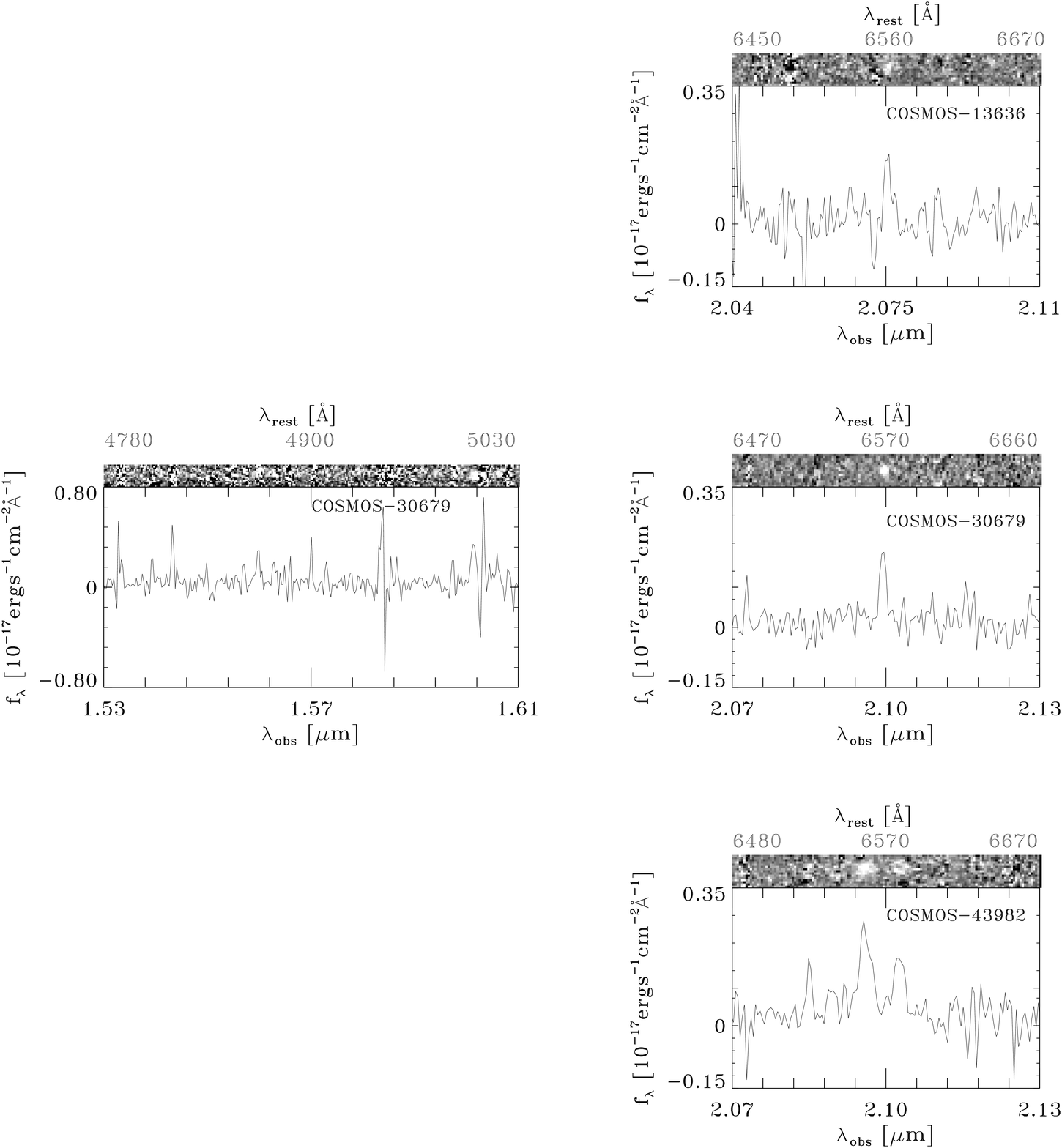}
\caption[]
{
Reduced 1D+2D spectra of 
COSMOS-13636 (top), COSMOS-30679 (middle),
and COSMOS-43982 (bottom).
$H$-band data were taken only for COSMOS-30679.
}
\label{fig:cosmos_spec}
\end{figure}

We reduced the NIRSPEC data using mainly IRAF tasks.
Details of the reduction procedure is described
in \cite{nakajima2012b}.
The flux-calibrated 1D spectra of the three objects are shown in
Figure \ref{fig:cosmos_spec}.

\subsection{Emission Line Detections and Measurements} \label{subsec:emission_confirments_and_mesurements}

We determine a line to be detected, if there exists an emission
line above the $3\sigma$ sky noise around the wavelength expected
from the Ly$\alpha$ redshift, where sky noise is calculated from
the spectrum within 100 \AA\ from the line wavelength.
Table \ref{tab:line_prop} summarizes the line detections.
For each detected line, we fit a Gaussian function to calculate
its central wavelength, FWHM, and flux.
The central wavelength is then converted into the vacuum wavelength
after correction for the heliocentric motion of the earth
\footnote{http://fuse.pha.jhu.edu/support/tools/vlsr.html},
and is adopted for the systemic redshift of the object.
For the four objects with multiple line detections,
we confirm that the redshifts of the lines agree within
$1\sigma$ errors, and adopt their weighted mean
for the systemic redshift.
{\sc [Oii]} is not used in the redshift measurement
because the 3726\AA/3729\AA\ doublet is not resolved in our spectra.
The H$\beta$ line for CDFS-3865 is not used, either,
because of the low $S/N$ ratio.
We also derive $1\sigma$ upper limits of fluxes for all undetected lines.
The emission line measurements are summarized in Table \ref{tab:line_prop}.

\begin{deluxetable*} {cccccccc}
\tablecolumns{7}
\tablewidth{0pt}
\tablecaption{Emission Line Properties \label{tab:line_prop}}
\tablehead{
\colhead{Object} & \colhead{Line} & \colhead{$\lambda_{\rm rest}$} & \colhead{$\lambda_{\rm obs}$} &  \colhead{$\lambda_{\rm corr.}$} & \colhead{z} & \colhead{$S/N$} & \colhead{Insuturment}\\
\colhead{} & \colhead{} & \colhead{(\AA)} & \colhead{(\AA)} &  \colhead{(\AA)} & \colhead{} & \colhead{} & \colhead{}\\
\colhead{(1)} & \colhead{(2)} & \colhead{(3)} & \colhead{(4)} &  \colhead{(5)} & \colhead{(6)} & \colhead{(7)} & \colhead{(8)}
}
\startdata
$$ &{\sc [Oii]} & 3727.00   & -$^a$                       & -$^a$       &  -$^a$   & $6.2$ & NIRSPEC\\
$$ & H$\beta$ & 4861.33 & $15418.9\pm 4.0$ & 15418.3 & $2.17162\pm0.00083$ & $3.2$ & MMIRS\\
$$ & {\sc [Oiii]} & 4958.91 & $15735.8\pm3.5$ & 15735.1 & $2.17310\pm0.00071$ & $5.4$ &MMIRS \\
CDFS-3865 & {\sc [Oiii]} & 5006.84 & $15887.1 \pm 1.5$ & 15886.4 & $2.17294 \pm 0.00030$ & $15.3$ & MMIRS \\
$$ & H$\alpha$ & 6562.85 & $20825.8\pm1.7$ & 20824.9 & $2.17315 \pm 0.00026$ & $16.0$ & MMIRS \\
$$ & {\sc [Nii]} & 6583.45 &  -  &  -  &  -  & $<$ $1$ & MMIRS \\
\hline \\
$$ & H$\beta$ & 4861.33 &  -  &  -  &  -  & $<$ $1$ & MMIRS  \\
$$ & {\sc[Oiii]} & 4958.91  &  -  &  -  &  -  & $<$ $1$ & MMIRS  \\
CDFS-6482 & {\sc[Oiii]} & 5006.84 & $16048.3 \pm 3.8$ & 16047.63 & $2.205144 \pm 0.00076$  & $8.8$ & MMIRS \\
$$ & H$\alpha$ & 6562.85 & $21036.9\pm1.9$ & 21036.02 & $2.205317 \pm 0.00029$ & $5.5$ & MMIRS \\
$$ & {\sc[Nii]}& 6583.45 &  -  &  -  &  -  & $<$ $1$ &MMIRS  \\
 \hline \\
COSMOS-13636 & H$\alpha$ & 6562.85 & $20753.5 \pm 1.0$ & $20752.36$ & $2.16210 \pm 0.00015$ & $7.1$ & NIRSPEC \\
$$ & {\sc[Nii]} & 6583.45 &  -  &  -  &  -  & $<$ $1$ & NIRSPEC\\
 \hline \\
$$ &{\sc [Oii]} & 3727.00   & -$^a$                       & -$^a$       &  -$^a$   & $4.0$ & NIRSPEC\\
$$ & H$\beta$ & 4861.33 &  -  &  -  &  -  & - $^{b}$ & NIRSPEC\\
$$ & {\sc[Oiii]} & 4958.91 &  -  &  -  &  -  &  - $^{b}$ & NIRSPEC\\
COSMOS-30679 & {\sc[Oiii]}& 5006.84 & $16013.8 \pm 1.2$ & 16013.62 & $2.19835 \pm 0.00024$ & $10.0$ & NIRSPEC \\
$$ & H$\alpha$ & 6562.85 & $20993.7 \pm 1.3$ & $20993.48$ & $2.19884 \pm 0.00019$ & $11.5$ & NIRSPEC\\
$$  &  {\sc[Nii]}   &   6583.45 &  -  &  -  &  -  & $<$$1$ &NIRSPEC\\
 \hline \\

COSMOS-43982   &   H$\alpha$    &   6562.85  & $20958.50\pm1.2$ &  20958.26 &  $2.19347\pm0.00018$ & $11.4$ & NIRSPEC\\
$$  &   {\sc [Nii]}    &   6583.45 &  $21026.20\pm2.1$ &  21025.96  & $2.19376\pm0.00032$ & $7.1$ & NIRSPEC
\enddata
\tablecomments{
The weighted mean redshifts for objects with multiple line detection
are $2.17306\pm0.00019$ (CDFS-3865),
$2.20530\pm0.00027$ (CDFS-6482), $2.19865\pm0.00015$ (COSMOS-30679), 
and $2.19354\pm0.00016$ (COSMOS-43982).
The symbol ``-'' indicates no detection.
(1) Object ID;
(2), (3) Line name and its rest-frame wavelength;
(4) Observed wavelength of the line;
(5) Wavelength of the line corrected for the LSR motion; 
(6) Redshift;
(7) Signal to noise ratio of the line detection; 
(8) Instrument.
}

\tablenotetext{a}{
{\sc [Oii]} redshift is not shown because it is not
reliably measured.
}
\tablenotetext{b}{
$S/N$ upper limit is not shown because the line is
contaminated by a strong OH line.
}

\end{deluxetable*}

\subsection{Checking the presence of AGNs} \label{subsec:agn_activity}

We examine whether our objects host an AGN in three ways.
First, we compare the sky coordinates of the objects with
those in deep archival X-ray and radio catalogues. 
For CDFS, we refer to the X-ray catalogue given by \cite{luo2008}, 
whose sensitivity limits are $1.9\times10^{-17}$ and $1.3\times10^{-16}$ 
ergs cm$^{-2}$ s$^{-1}$ for the $0.5-2.0$ and $2-8$ keV bands, respectively.
\footnote{CDFS   http://www2.astro.psu.edu/users/niel/cdfs/luo2008-chandra-catalog/paper001-cdfs-luo-table2.txt}
For COSMOS, we use the X-ray catalogue by \cite{elvis2009}, 
whose sensitivity limits are $1.9\times10^{-16}$ ($0.5-2.0$ keV band),  
$7.3\times10^{-16}$ ($2-10$ keV band), 
and $5.7\times10^{-16}$ ergs cm$^{-2}$ s$^{-1}$ ($0.5-10$ keV band). 
We also refer to the radio catalogue constructed by \cite{schinnerer2010}, 
whose sensitivity limits are $10-40\mu$ Jy/beam.
\footnote{COSMOS http://irsa.ipac.caltech.edu/data/COSMOS/}
No counterpart for the LAEs is found in any of the catalogues.

Second, we look for three high ionization state lines typical of AGNs,
C{\small IV}$\lambda 1549$, He{\small II}$\lambda 1640$,
and C{\small III]}$\lambda 1909$, in the spectra,
and detect none of them.

\begin{figure}[h]
\includegraphics[scale=0.8]{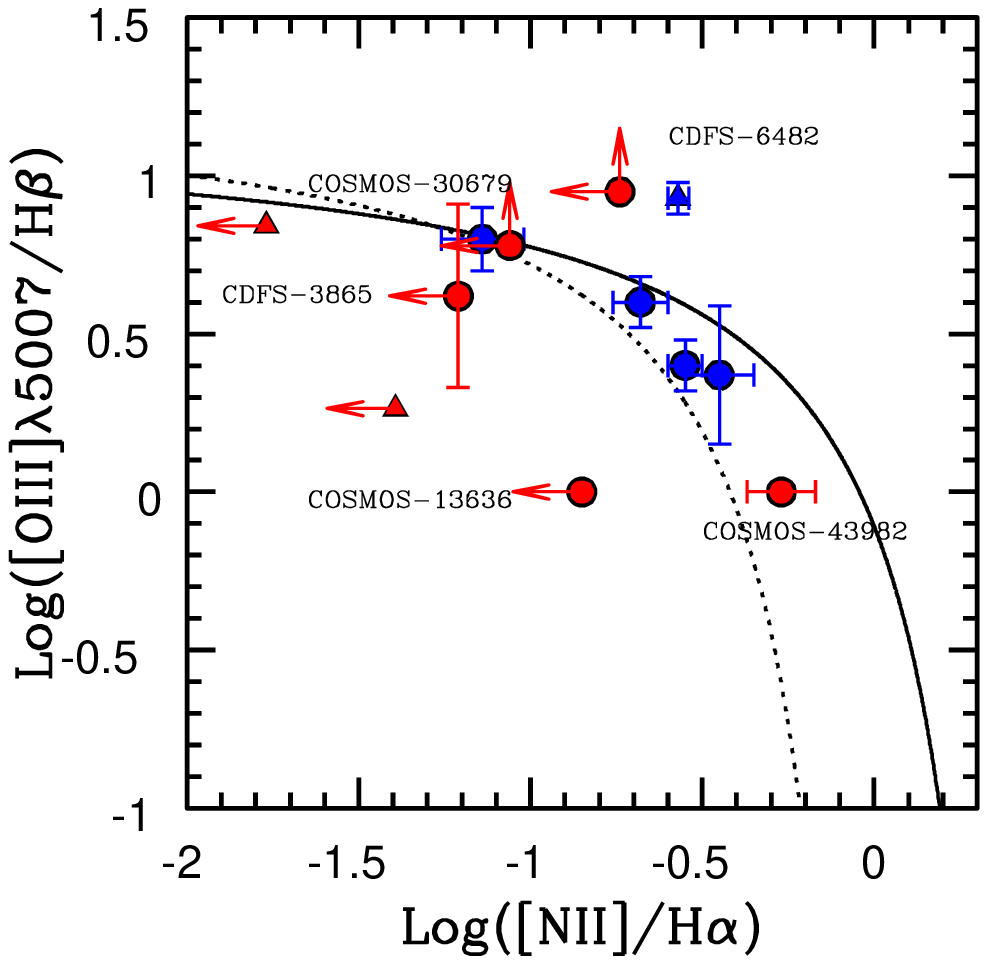}
\caption[]
{
BPT diagram. The solid curve represents
the boundary between star-forming galaxies and AGNs 
proposed by \cite{kewley2001}, while the dotted curve is the boundary
empirically defined using SDSS objects \citep{kauffmann2003}.
The red circles denote our LAEs.
For the purpose of display, COSMOS-13636 and COSMOS-43982, whose
[O{\small III}] / H$\beta$ is not constrained, are placed at
log([O{\small III}] / H$\beta$) = 0.0.
The red triangles, blue circles, and blue triangle 
show, respectively, 
two $z\sim2.3$ LAEs in \citet{finkelstein2011b}, 
four bins of $z\sim2.2$ LBGs in \citet{erb2006a}, 
and a lensed LBG at $z=2.73$ \citep{finkelstein2009}.
}
\label{fig:bpt}
\end{figure}

Finally and most importantly,
we apply the BPT diagnostic diagram \citep{baldwin1981}
to our objects, as shown in Figure \ref{fig:bpt}.
The solid curve in Figure \ref{fig:bpt} shows the boundary between
star-forming galaxies and AGNs proposed by \cite{kewley2001}
using photoionization models,
while the dotted curve is the boundary
empirically defined by SDSS objects \citep{kauffmann2003}.
In both cases, star-forming galaxies fall below the curve.
None of our objects 
has detections of all four lines necessary for the BPT diagram.
Thus, as shown in Figure \ref{fig:bpt}, any object but one 
logically has a possibility of being in the AGN regime.
However, we conclude from the following discussion
that all but COSMOS-43982 are star-forming galaxies.
CDFS-3865 falls below the curves, indicating this is a starburst galaxy.
CDFS-6482 has a relatively high [O{\small III}] / H$\beta$
value (lower limit), but [N{\small II}]/H$\alpha$ is not
so high.
This high [O{\small III}] / H$\beta$ ratio may not be due to
the presence of AGN but due to a higher gas temperature of
the H{\small II} regions, as has been pointed out for some
high-redshift star-forming galaxies \citep{erb2006a}.
Indeed, there are few local AGNs distributed in the 
range of log([N{\small II}]/H$\alpha$) $\lesssim$ \ $-0.5$ 
(e.g., Figure 3 in \citealt{finkelstein2011b}).

\cite{finkelstein2009} have also classified a similar object to ours
as a star-forming galaxy (blue triangle in Figure \ref{fig:bpt}).
A similar argument is made for COSMOS-13636 and
COSMOS-30679 whose [N{\small II}]/H$\alpha$ ratios are also modest.
In contrast, COSMOS-43982 has a very high [N{\small II}]/H$\alpha$
ratio. Because we infer that this object is more likely to be an AGN
than a star-forming galaxy,
we do not use this object in the following discussion.

\section{SED Fitting}\label{sec:sed}

We perform SED fitting to our objects to derive
dust extinction values, SFRs, and stellar masses.
The procedure of the SED fitting is the same as
that of \cite{ono2010a}.
For the CDFS objects, we use 12 bandpasses:
$\it{B, V, R, I, z, J, H, K}$ data from the MUSYC public data release
\footnote{http://www.astro.yale.edu/MUSYC/
} \citep{cardamone2010},
and $\it{Spitzer}$/IRAC 3.6, 4.5, 5.8, and 8.0 $\mu$m photometry
from the $\it{Spitzer}$ legacy survey of the UDS field.
For the COSMOS objects, we use 11 bandpasses:
$B, V, r', i'$, and $z'$ data
taken with Subaru/Suprime-Cam,
$\it{J}$ data taken with UKIRT/WFCAM,
$\it{K_s}$ data taken with CFHT/WIRCAM \citep{mccracken2010},
and $\it{Spitzer}$/IRAC 3.6, 4.5, 5.8, and 8.0 $\mu$m photometry
from the $\it{Spitzer}$ legacy survey of the UDS field.
We use neither ${\it u^{*}/U}$ nor NB387-band data,
since the photometry of these bands is contaminated
by the IGM absorption and/or Ly$\alpha$ emission.
Tables \ref{tab:photometry_cdfs} and \ref{tab:photometry_cosmos} summarize
the broadband photometry of our objects.
Basically, the uncertainties in the optical photometry include
both photometric errors and systematic errors associated with aperture
correction and photometric calibration.

\begin{figure*}[!t]
\begin{center}
\includegraphics[scale=0.57]{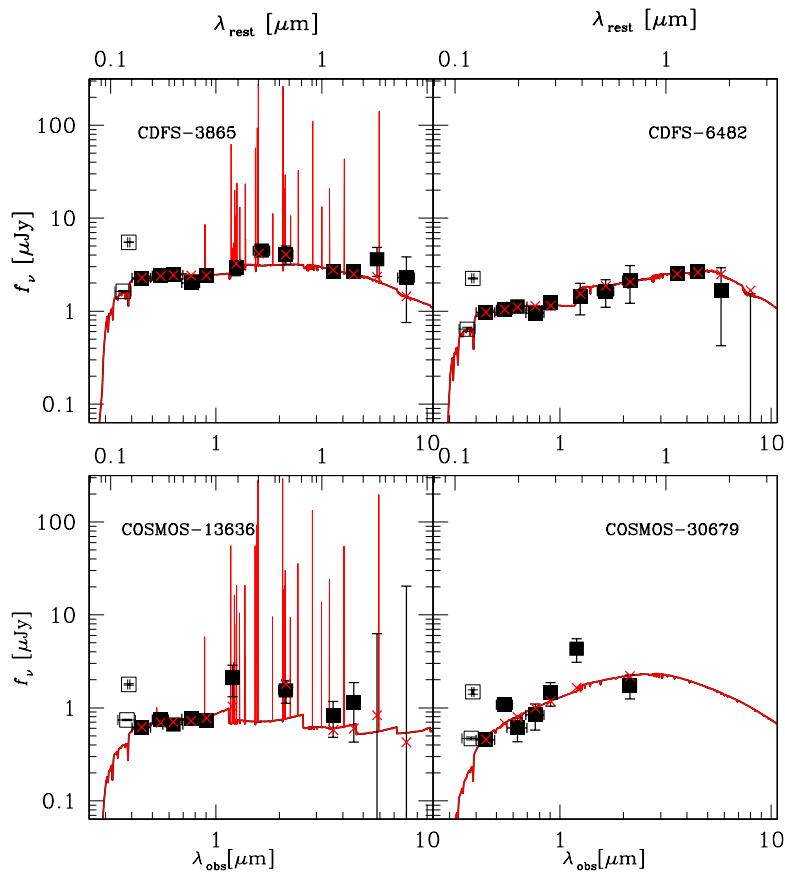}
\end{center}
\caption[]
{
Upper (lower) panel shows the SEDs of CDFS (COSMOS) objects.
The filled squares denote the photometry points used for
SED fitting, while the open squares are those
omitted in SED fitting due to the contamination
of Ly$\alpha$ emission and IGM absorption.
The red lines present the best-fit model
spectra, and the red crosses correspond to
the flux densities at individual passbands 
expected from the best-fit models.
}
\label{fig:sed_all}

\vspace{20pt}

\end{figure*}

We use the stellar population synthesis model of GALAXEV
\citep{bc03} including nebular emission
\citep{schaerer_de_barros2009}, and 
adopt a Salpeter initial mass function \citep{salpeter1955}.
Because LAEs are metal poor star-forming galaxies,
we choose constant star formation models with
a metallicity of $Z/Z_\odot=0.2$.
We use Calzetti's law \citep{calzetti2000}
for the stellar continuum extinction, $E(B-V)_{*}$.
We apply \cite{madau1995}'s prescription 
to correct for the IGM attenuation;
at $z\simeq2.2$, continuum photons shortward of Ly$\alpha$
are absorbed by 18 $\%$.
Figure \ref{fig:sed_all} shows the best-fit model spectra
with the observed flux densities for individual objects.
The physical quantities derived from SED fitting are summarized 
in Table \ref{tab:prop_sed}.
Because the continuum emission of COSMOS-30679 is 
blended with a foreground object, we deblend the
source with GALFIT (\citealt{peng2002}; 
see \citealt{nakajima2012b} for details).
While photometry data both before and after deblending 
are given in Table \ref{tab:photometry_cosmos}, 
we use the latter for SED fitting.

\begin{deluxetable*}{ccccccccccccc}
\tablecolumns{13}
\tablewidth{0pt}
\tablecaption{Broadband Photometry of our Sample (CDFS) \label{tab:photometry_cdfs}}
\tablehead{
\colhead{Object} & \colhead{\it{B} } & \colhead{\it{V}} & \colhead{\it{R}} & \colhead{\it{I}}
& \colhead{\it{z}} & \colhead{\it{J}} & \colhead{\it{H}} & \colhead{\it{K}} & \colhead{[3.6]}
& \colhead{[4.5]} & \colhead{[5.8]} & \colhead{[8.0]}
}
\startdata
CDFS-3865 & 23.01 & 22.94 & 22.92 & 23.14 & 22.93 & 22.73 & 22.27 & 22.38 & 22.82 & 22.82 & 22.51& 23.00\\
$$& (0.01) & (0.01) & (0.01) & (0.07) & (0.09) & (0.18) & (0.12) & (0.23) & (0.05) & (0.08) & (0.32) & (0.56)\\
\hline
CDFS-6482 & 23.93 & 23.87 & 23.78 & 23.95 & 23.67 & 23.50 & 23.36 & 23.07 & 22.88 & 22.83 & 23.34& 99.99\\
$$& (0.02) & (0.03) & (0.03) & (0.14) & (0.16) & (0.34) & (0.31) & (0.39) & (0.05) & (0.08) & (0.60) & (-)
\enddata
\tablecomments{
All magnitudes are total magnitudes.
$99.99$ mag indicates a negative flux density.
Magnitudes in parentheses are $1\sigma$ uncertainties.
}

\end{deluxetable*}

\begin{deluxetable*}{cccccccccccc}
\tablecolumns{12}
\tablewidth{0pt}
\tablecaption{Broadband Photometry of our Sample (COSMOS) \label{tab:photometry_cosmos}}
\tablehead{
\colhead{Object} & \colhead{\it{B}} & \colhead{\it{V}} & \colhead{$r'$} & \colhead{$i'$}
& \colhead{$z'$} & \colhead{\it{J}} & \colhead{\it{K$_{s}$}} & \colhead{[3.6]}
& \colhead{[4.5]} & \colhead{[5.8]} & \colhead{[8.0]}
}
\startdata
COSMOS-13636 & 24.43 & 24.21 & 24.35 & 24.19 & 24.24 & 23.10 & 23.43 & 24.10 & 23.75 &  99.99 & 99.99\\
$$ & (0.01) & (0.03) & (0.03) & (0.04) & (0.09) & (0.34) & (0.26) & (0.38) & (0.53) & (-) & (-)\\
\hline
COSMOS-30679$^{a}$ & 24.05 & 23.12 & 22.91 & 22.46 & 22.33 & 21.15 & 21.83 & 22.12 & 22.57 &  99.99 & 23.06\\
$$ & (0.01) & (0.01) & (0.01) & (0.01) & (0.02) & (0.07) & (0.07) & (0.07) & (0.21) & (-) & (2.54)\\
\\
COSMOS-30679$^{b}$ & 24.76 & 23.82 & 24.44 & 24.09 & 23.49 & 22.31 & 23.29 & - & - &  - & -\\ 
$$ & (0.03) & (0.11) & (0.28) & (0.30) & (0.27) & (0.27) & (0.28) & ( - ) & ( - ) & ( - ) & ( - )
\enddata
\tablenotetext{a}{
Before deblending of a foreground source.
}
\tablenotetext{b}{
After deblending of a foreground source.
}

\enddata
\tablecomments{
All magnitudes are total magnitudes.
$99.99$ mag indicates a negative flux density.
Magnitudes in parentheses are $1\sigma$ uncertainties.
}

\end{deluxetable*}

\begin{deluxetable}{ccccc}
\tablecolumns{5}
\tablewidth{0pt}
\tablecaption{Results of SED fitting \label{tab:prop_sed}}
\tablehead{
\colhead{Object} & \colhead{$E(B-V)_{*}$} & \colhead{$SFR_{SED}$} & \colhead{$M_{*}$} & \colhead{$\chi^{2}$}\\
\colhead{} & \colhead{} & \colhead{($M_\odot$ yr$^{-1}$)} & \colhead{($10^{9}$ $M_\odot$)} & \colhead{}
\\
\colhead{(1)} & \colhead{(2)} & \colhead{(3)} & \colhead{(4)} & \colhead{(5)}
}
\startdata
CDFS-3865 & $0.185^{+0.009}_{-0.009}$ & $312^{+41}_{-43}$  &  $3.18^{+0.21}_{-0.13}$ & $17$\\
CDFS-6482 &  $0.185^{+0.026}_{-0.018}$ & $83^{+34}_{-18}$ & $5.30^{+1.18}_{-0.80}$ & $5$\\
COSMOS-13636 &  $0.273^{+0.018}_{-0.079}$ & $1311^{+17574}_{-1244}$ & $1.99^{+0.39}_{-1.06}$ & $21$\\
COSMOS-30679 &  $0.528^{+0.026}_{-0.026}$ & $7510^{+201248}_{-3097}$ & $19.75^{+6.53}_{-5.80}$ &$20$
\enddata
\tablenotetext{}{Notes.
Stellar metallicity is fixed to 0.2 $Z_{\odot}$.\\
(1) Object ID;
(2) Dust extinction;
(3) Star formation rate;
(4) Stellar mass; 
(5) $\chi^{2}$ of the fitting.
}
\end{deluxetable}

\section{Physical Quantities of Our LAEs} \label{sec:physical_quantities}

\subsection{Velocity Offset Between Ly$\alpha$ and Nebular Lines}
\label{subsec:velocity_offset}

We calculate a quantity called the velocity offset of
the Ly$\alpha$ line:
\begin{equation}
\Delta v_{\rm Ly\alpha}
= c \frac{z_{{\rm Ly}\alpha} - z_{\rm sys}}{1+z_{\rm sys}},
\end{equation}
where $z_{{\rm Ly}\alpha}$ is the redshift of Ly$\alpha$ emission
and $z_{\rm sys}$ is the systemic redshift
given in Table \ref{tab:line_prop}.

\begin{figure*}
\begin{center}
\includegraphics[scale=0.6]{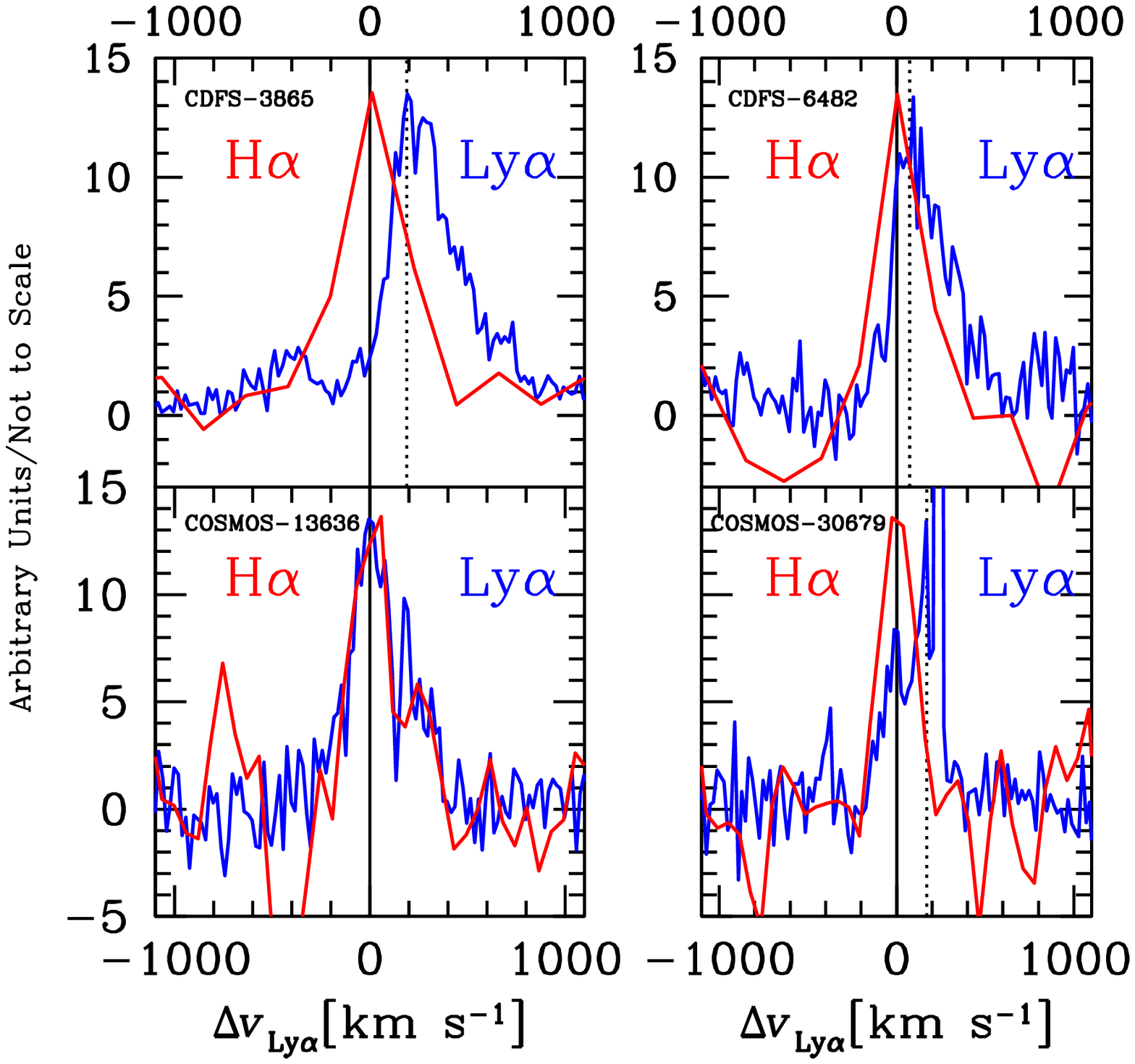}
\end{center}
\caption[]
{
Comparison between Ly$\alpha$ (blue line) and H$\alpha$ (red line) 
emission profiles.
The solid and dotted lines indicate 
the systemic and Ly$\alpha$ redshifts,
respectively.%
{\it Top left:} CDFS-3865 with
$\Delta v_{\rm Ly\alpha}$ = $190^{+99}_{-18}$ km s$^{-1}$.
{\it Top right:} CDFS-6482 with
$\Delta v_{\rm Ly\alpha}$ = $75^{+52}_{-25}$ km s$^{-1}$.
{\it Bottom left:} COSMOS-13636 with
$\Delta v_{\rm Ly\alpha}$  = $18 \pm 16$ km s$^{-1}$.
{\it Bottom right:} COSMOS-30679 with
$\Delta v_{\rm Ly\alpha}$  = $170 \pm 16$ km s$^{-1}$.
}
\label{fig:velocity_offset_all}
\end{figure*}

We show the Ly$\alpha$ and H$\alpha$ profiles 
in Figure \ref{fig:velocity_offset_all}.
Most objects have a Ly$\alpha$ profile which is asymmetric
and/or double peaked.
Similar to previous studies \citep{steidel2010,yang2011},
we define $z_{{\rm Ly}\alpha}$ to be the redshift
corresponding to the wavelength at the highest peak.
The peak wavelength is determined as follows.
First, we roughly constrain the peak position by the eye.
For the CDFS objects, we cannot identify where is the highest
peak, since the second peak is within the $1\sigma$ error
in the height of the first peak.
In these cases, we regard both as peak candidates
and include their wavelength difference as the error
in $z_{{\rm Ly}\alpha}$
(as listed in Table \ref{tab:observation}).
Then, we fit a Gaussian to the profile only around the peak
to derive the peak wavelength,
in order to avoid systematic effects due to the asymmetric
profile.
The Ly$\alpha$ line of COSMOS-30679 is partly contaminated
by a cosmic ray.
It is, however, unlikely that there is a true peak
under the cosmic ray.
Even if there is a flux peak just under the cosmic ray,
our discussion remains unchanged.

We obtain $\Delta v_{\rm Ly\alpha}$ = $190^{+99}_{-18}$ km s$^{-1}$
(CDFS-3865),
$75^{+52}_{-25}$ km s$^{-1}$ (CDFS-6482),
$18\pm16$ km s$^{-1}$ (COSMOS-13636),
and $170\pm16$ km s$^{-1}$ (COSMOS-30679).
Thus, three out of the four have a positive $\Delta v_{\rm Ly\alpha}$
beyond the $2\sigma$ uncertainty.

\begin{figure}
\begin{center}
\includegraphics[scale=0.42]{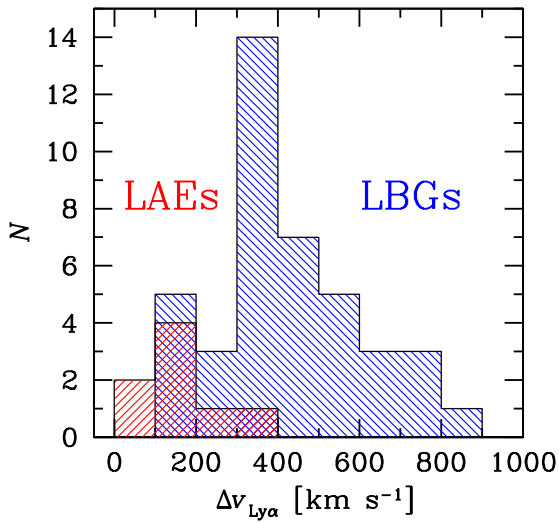}
\end{center}
 \caption[]
{
Histograms of $\Delta v_{\rm Ly\alpha}$ 
for the eight LAEs studied in this paper (red) and 
41 LBGs given by \cite{steidel2010} (blue).
}
\label{fig:histogram}
\end{figure}

There are four LAEs in the literature which have both
a Ly$\alpha$ spectrum and a systemic-velocity measurement
from a nebular line:
two from \cite{mclinden2011} and two from \cite{finkelstein2011b}.
The $\Delta v_{\rm Ly\alpha}$ of these four LAEs ranges
from $\sim 100$ to $\sim 300$ km s$^{-1}$,
which is similar to our $\Delta v_{\rm Ly\alpha}$ measurements.
We combine these four LAEs with our four to construct
a sample of eight LAEs at $z \sim 2-3$,
and investigate average physical properties of LAEs
using them.
Figure \ref{fig:histogram} presents the histogram of
$\Delta v_{\rm Ly\alpha}$ for the eight LAEs.
All eight have a positive but relatively small 
$\Delta v_{\rm Ly\alpha}$ of up to $\sim 300$ km s$^{-1}$.
The average of the eight is 
$\Delta v_{\rm Ly\alpha}= 175 \pm 35$ km s$^{-1}$, 
which is systematically smaller, by $\sim 200-300 $ km s$^{-1}$,
than that of LBGs,
$\Delta v_{\rm Ly\alpha} \simeq 400$ km s$^{-1}$ 
\citep{pettini2002,shapley2003,steidel2010,rakic2011}, 
at similar redshifts of $z \sim 2$--$3$. 
There are also two Ly$\alpha$ blobs 
(LABs; e.g., \citealt{steidel2000,matsuda2004,yang2009})
at $z\sim 2.3$ whose $\Delta v_{\rm Ly\alpha}$ is measured \citep{yang2011}. 
Both have $\Delta v_{\rm Ly\alpha}$= 0 -- 200 km s$^{-1}$ 
which is comparable with our measurements for LAEs 
but smaller than those for LBGs.

\begin{figure}
\begin{center}
\includegraphics[width=9.3cm]{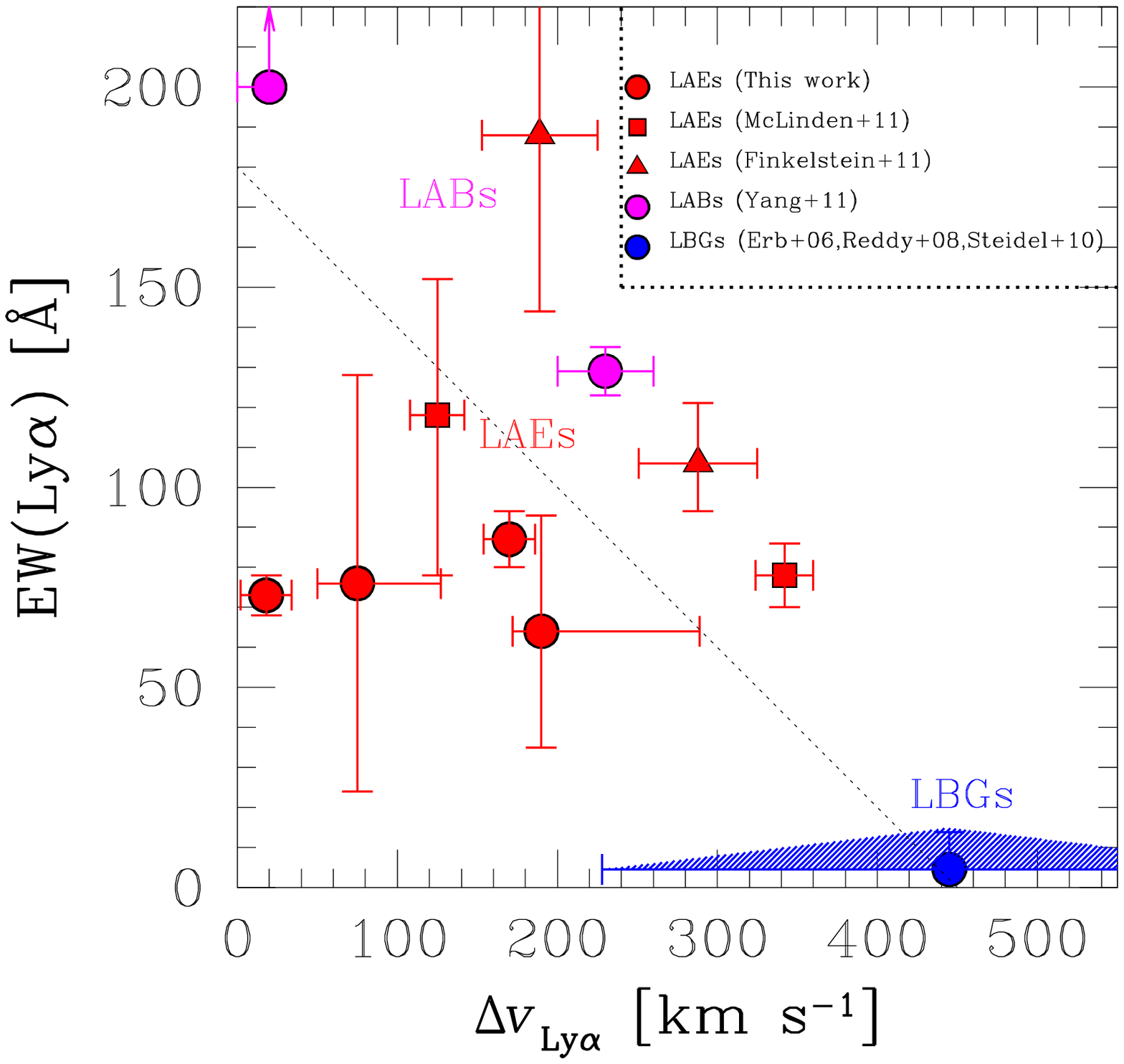}
\end{center}
\caption[]
{
Rest-frame EW(Ly$\alpha$) plotted against $\Delta v_{\rm Ly\alpha}$.
The red circles are our LAEs. 
The red square and the red triangle show the LAEs by 
\cite{mclinden2011} and \cite{finkelstein2011b}, respectively.
The blue symbol indicates the average of 41 LBGs, 
with the error bars corresponding to the 68 percentiles 
of the $\Delta v_{\rm Ly\alpha}$ distribution \citep{steidel2010} 
and the EW distribution \citep{reddy2008}. 
The magenta circles denote the LABs by \cite{yang2011}.
The dotted line is the best-fit linear function to all the data points.
}
\label{fig:dv_ew}
\end{figure}

Figure \ref{fig:dv_ew} shows the rest-frame Ly$\alpha$ EW 
as a function of $\Delta v_{\rm Ly\alpha}$,
where data for LBGs  \citep{reddy2008,steidel2010} and 
LABs \citep{yang2011} are also included
in order to cover a wide baseline of EW, $0-200$ \AA. 
We fit a linear function to the data points
(dotted line in Figure \ref{fig:dv_ew}), and find
an anti-correlation between $\Delta v_{\rm Ly\alpha}$
and Ly$\alpha$ EW that EW(Ly$\alpha$) decreases with 
increasing $\Delta v_{\rm Ly\alpha}$.
It is known that there is an anti-correlation between 
$\Delta v_{\rm Ly\alpha}$ - $\Delta v_{\rm abs}$ and EW(Ly$\alpha$) 
in LBGs at $z\sim3$ (\citealt{shapley2003}). 
However, it should be noted that our finding is different 
from that, 
because we measure the velocity offset of Ly$\alpha$ directly 
from the systemic velocity, 
while $\Delta v_{\rm Ly\alpha}$ - $\Delta v_{\rm abs}$ 
depends on the kinematics of gas outflows and 
the Ly$\alpha$ emission mechanism.

\subsection{Velocity Offset Between LIS Absorption Lines And Nebular Lines} 
\label{subsec:metal_lines}

We examine LIS absorption lines 
using Magellan/MagE echellette spectrograph data
with a medium-high spectral resolution of $R\sim 4100$
(M. Rauch et al., in preparation). 
We inspect one order of the echelle spectra ranging
from  3788 \AA \ to 4419 \AA \ in the observed frame.
This wavelength range covers
Si {\small II} $\lambda 1260$,
O {\small I} $\lambda 1302$,
Si {\small II} $\lambda 1304$,
and C {\small II} $\lambda 1335$ lines.
These LIS absorption lines 
are associated with the neutral interstellar medium 
(ISM: \citealt{pettini2002,shapley2003}).

Since none of these lines is identified in the individual spectra 
because of the too faint continua, 
we make a composite spectrum of the four objects as follows.
For each object, 
we shift individual spectral data from the observed 
to the rest frame using $z_{\rm sys}$ given in 
Table \ref{tab:line_prop}. 
Then, we stack the spectra of the four objects 
with statistical weights based on the $S/N$ ratios 
at 1250--1340 \AA.

\begin{figure}
\begin{center}
\includegraphics[width=6cm]{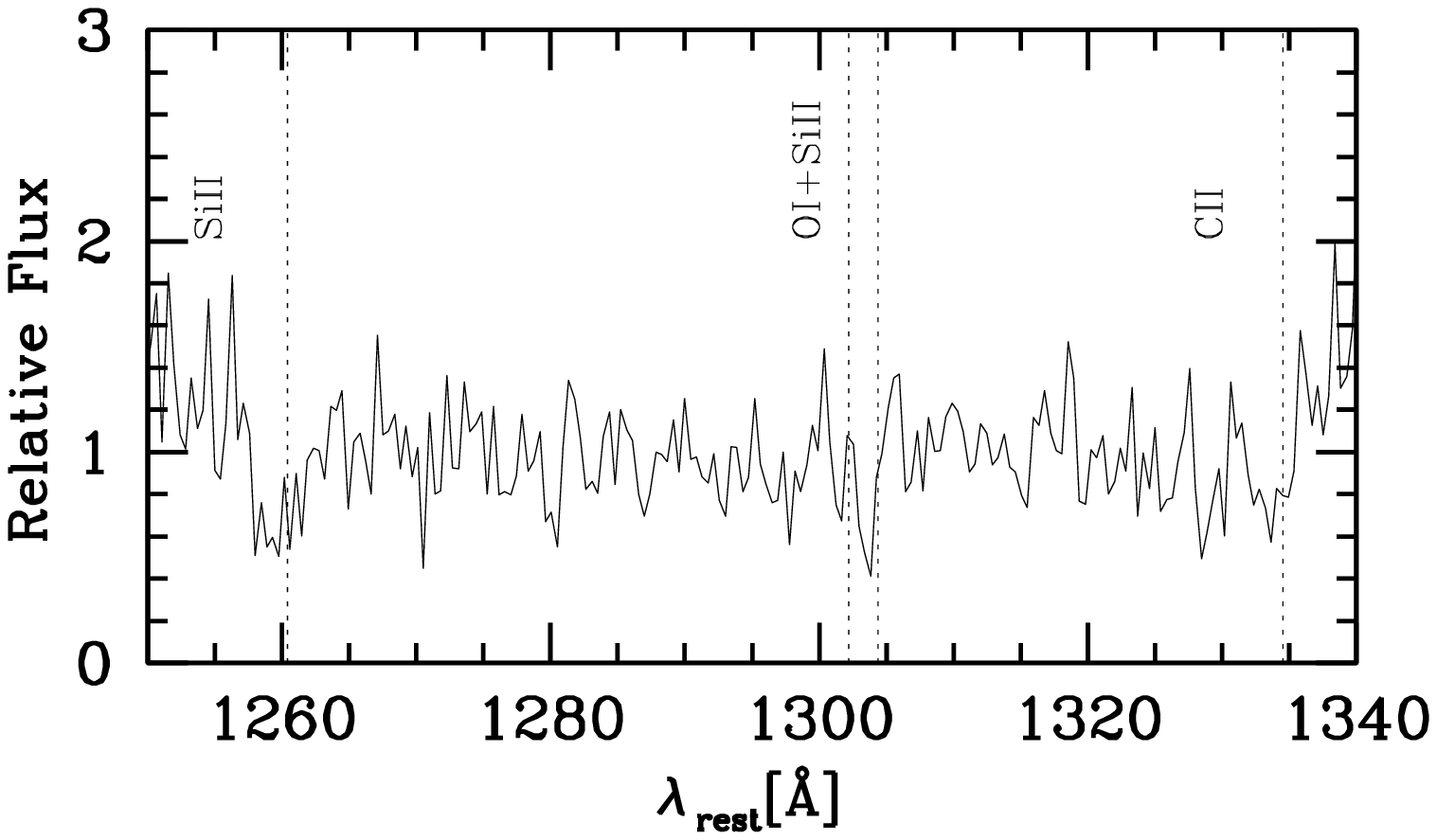}
\end{center}
\caption[]
{
Composite FUV spectrum of the four LAEs.
The spectrum has been normalized to unity in the continuum.
The dotted lines indicate the rest-frame vacuum
wavelengths of four LIS absorption lines.
The spectrum is plotted with 0.5 \AA\ pix$^{-1}$ sampling 
so that the profile of the blended O {\sc i} and  Si {\sc ii} lines 
be clearly shown.}
\label{fig:LIS}
\end{figure}

Figure \ref{fig:LIS} presents the composite spectrum. 
An absorption feature is seen 
near each of the four lines' wavelengths.
For each feature, we fit a Gaussian profile
to obtain the rest-frame equivalent width, EW(LIS), 
and the velocity offset of the line center 
with respect to the systemic velocity, $\Delta v_{\rm abs}$, 
as summarized in Table \ref{tab:absorption}. 
We securely detect the Si {\small II} $\lambda 1260$ line and
the blended O {\small I} $\lambda 1302$ + Si {\small II} $\lambda 1304$
lines at the 5.2 $\sigma$ and 3.7 $\sigma$ levels, respectively, 
and marginally detect the 
C {\small II} $\lambda 1335$ line
at the 2.0$\sigma$ level.
Note that all lines are blue-shifted with respect to 
the systemic velocity.

The weighted mean offset velocity of
these absorption lines is
$\Delta v_{\rm abs} = -179 \pm 73$ km s$^{-1}$.
This value is comparable with those of LBGs, which are 
typically 
$\Delta v_{\rm abs} \sim -150$ km s$^{-1}$
(\citealt{shapley2003,steidel2010}).
Thus, LAEs and LBGs have similar $\Delta v_{\rm abs}$ values, 
in contrast to the significant difference in
$\Delta v_{\rm Ly\alpha}$.

\begin{deluxetable}{cccccc}
\tablecolumns{6}
\tablewidth{0pt}
\tablecaption{Interstellar Absorption Features \label{tab:absorption}}
\tablehead{
\colhead{Ion} & \colhead{$\lambda_{\rm rest}$} & \colhead{$f$} & \colhead{EW(LIS)} & \colhead{$\sigma$} & \colhead{$\Delta v_{\rm abs}$}\\
\colhead{} & \colhead{(\AA)} & \colhead{} & \colhead{(\AA)} & \colhead{(\AA)} & \colhead{(km s$^{-1}$)}
\\
\colhead{(1)} & \colhead{(2)} & \colhead{(3)} & \colhead{(4)} & \colhead{(5)} & \colhead{(6)}
}
\startdata
Si {\sc ii} & 1260.4221& 1.007   & $-1.21$        &0.23         & $-162 \pm 95$ \\
O {\sc i}   & 1302.1685& 0.04887 & $-1.02^{(7)}$  &0.28$^{(7)}$ & $-174 \pm 180^{(7)} $\\
Si {\sc ii} & 1304.3702& 0.094   & $-1.02^{(7)}$  &0.28$^{(7)}$ & $-174 \pm 180^{(7)} $\\
C {\sc ii}  & 1334.5323& 0.1278  & $-0.52$        &0.25         & $-209 \pm 110 $ 
\enddata
\tablenotetext{}{
(1) Absorption line;
(2) Rest-frame vacuum wavelength;
(3) Transition oscillator strength (see e.g., \citealt{pettini2002,shapley2003});
(4)--(5) Rest-frame EW and its 1\ $\sigma$; 
(6) Velocity offset. 
(7) Values for the blended O {\sc i} $\lambda 1302$ 
and Si {\sc ii} $\lambda 1304$ lines assuming that 
the central rest-wavelength wavelength is
$\lambda$ = $1303.2694$ \AA.
}
\end{deluxetable}

\subsection{Dust Extinction} \label{subsec:dust}

The color excess of the stellar continuum 
obtained from the SED fitting is 
$E(B-V)_{*} = 0.185^{+0.009}_{-0.009}$ (CDFS-3865),
$0.185^{+0.026}_{-0.018}$ (CDFS-6482),
$0.273^{+0.018}_{-0.079}$ (COSMOS-13636), 
and $0.528^{+0.026}_{-0.026}$
(COSMOS-30679) for \cite{calzetti2000}'s extinction law.
These values are comparable to those of
relatively faint (stacked) LAEs at $z\sim 2$, 
$E(B-V)_{*}=0.27^{+0.01}_{-0.03}$ \citep{nakajima2012a} 
and  
$E(B-V)_{*} = 0.22^{+0.06}_{-0.13}$ \citep{guaita2011}, 
but are larger than those of the two bright LAEs
observed by \cite{finkelstein2011b}, 
$E(B-V)_{*} = 0.09\pm 0.05$ and $E(B-V)_{*} = 0.10\pm 0.10$.
Since our objects are as bright as 
\cite{finkelstein2011b}'s, 
this difference may indicate that LAEs at $z\sim 2$ have
a wide range of dust extinction
even at the same Ly$\alpha$ luminosity range.

Since the spectra of CDFS-3865 and CDFS-6482 cover H$\beta$
as well as H$\alpha$,
\footnote{COSMOS-30679 also has H$\beta$ data, but
the line is heavily contaminated by a sky emission line.},
we measure the Balmer decrement (H$\alpha$/H$\beta$) 
to constrain the color excess of nebular emission, $E(B-V)_{\rm gas}$.
The decrement values obtained are 
$2.96\pm0.87$ for CDFS-3865 
and $>1.74$ (using H$\beta$'s $2\sigma$ limit) for CDFS-6482, 
which are converted into $E(B-V)_{\rm gas}=0.03^{+0.27}_{-0.03}$ 
and $E(B-V)_{\rm gas} > 0$, respectively,
adopting the intrinsic ratio of 2.86 \citep{osterbrock1989}
and \cite{calzetti2000}'s extinction law.

\begin{figure}
\begin{center}
\includegraphics[width=7.2cm]{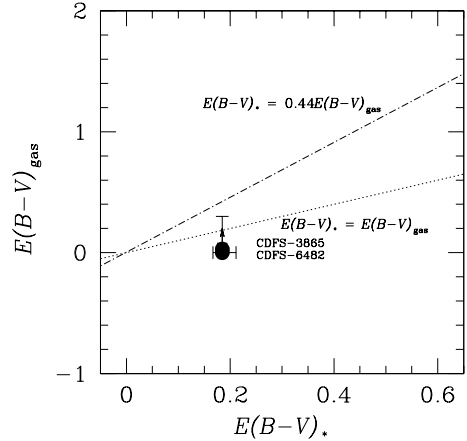}
\end{center}
\caption[]
{
$E(B-V)_{\rm gas}$ plotted against $E(B-V)_{*}$ for our two LAEs 
with a Balmer decrement measurement.
The dotted and dashed lines correspond to 
$E(B-V)_{*} = E(B-V)_{\rm gas}$ \citep{erb2006b} and
$E(B-V)_{*} = 0.44 E(B-V)_{\rm gas}$ \citep{calzetti2000}, 
respectively.
}
\label{fig:dust_dust}
\end{figure}

Figure \ref{fig:dust_dust} plots $E(B-V)_{*}$ vs. $E(B-V)_{\rm gas}$
for these objects.
CDFS-3865 has $E(B-V)_{\rm gas} \sim E(B-V)_{*}$.
Similarly, CDFS-6482's lower limit of $E(B-V)_{\rm gas}$
is consistent with its $E(B-V)_{*}$.
Thus, hereafter, we assume $E(B-V) \equiv 
E(B-V)_{\rm gas} \simeq  E(B-V)_{*}$ 
as proposed by \cite{erb2006b} for starburst galaxies.

\subsection{Star Formation Rate} \label{subsec:sfr}

The H$\alpha$ luminosity is thought to be a reliable indicator
of SFR.
We calculate the SFRs of our LAEs from their dust-corrected H$\alpha$
luminosities using \cite{kennicutt1998}'s formula:
\begin{equation}
SFR (M_{\odot}\ {\rm yr}^{-1})
 = 7.9 \times 10^{-42} L({\rm H}\alpha)\ ({\rm erg}\ {\rm s}^{-1}),
\end{equation}
where a Salpeter IMF is assumed.
We obtain $SFR = 190\pm13 M_{\odot}\ {\rm yr}^{-1}$ (CDFS-3865),
$48^{+10}_{-9}  M_{\odot}\ {\rm yr}^{-1}$ (CDFS-6482),
$17^{+3}_{-5}  M_{\odot}\ {\rm yr}^{-1}$ (COSMOS-13636),
and $45\pm5 M_{\odot}\ {\rm yr}^{-1}$ (COSMOS-30679).
These values are comparable to
those of \citet{finkelstein2011b}'s two LAEs,
$36.8\pm5.8\ M_{\odot}$ yr$^{-1}$ and $23.6\pm6.5\ M_{\odot}$ yr$^{-1}$,
derived from the H$\alpha$ luminosity, 
but are larger than 
the average values of $z\simeq 2.2$ LAEs obtained
from stacking analysis, $10-20\ M_{\odot}\ {\rm yr}^{-1}$
\citep{nilsson2011,nakajima2012a}.
This difference is reasonable, because LAEs with 
H$\alpha$ detection are generally brighter than average LAEs.

\begin{figure}
 \begin{center}
 \includegraphics[scale=0.35]{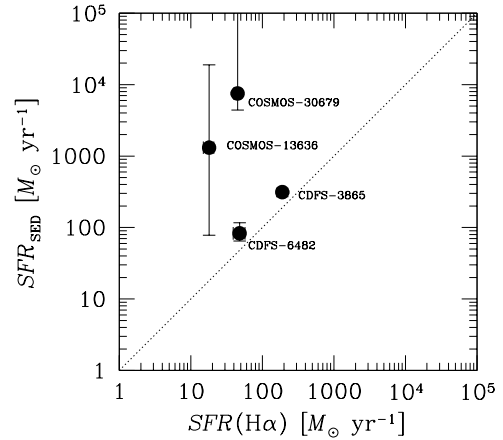}
\end{center}
 \caption[]
{
SFRs from SED fitting and from the H$\alpha$ luminosity.
From left to right, plotted are 
COSMOS-13636, COSMOS-30679, CDFS-6482,  and CDFS-3865.
}
\label{fig:sfr_sfr}
\end{figure}

Figure \ref{fig:sfr_sfr} plots the SFR from SED fitting 
against that from the H$\alpha$ luminosity for our objects.
Although the two SFR values 
are comparable for CDFS-3865 and CDFS-6482,
they are largely different for COSMOS-13636 and COSMOS-30679, 
indicating that 
models with very different star formation histories 
can fit the observed SED almost equally well.
This figure thus demonstrates that it is important to measure 
the H$\alpha$ luminosity to derive a reliable SFR.
In the rest of this paper, we quote the SFRs obtained from
the H$\alpha$ luminosity.

\subsection{Ly$\alpha$ Escape Fraction}\label{subsec:escape_fraction}
The Ly$\alpha$ escape fraction of a galaxy, $f^{{\rm Ly}\alpha}_{\rm esc}$,
is defined as the ratio of the observed Ly$\alpha$ flux 
to the intrinsic Ly$\alpha$ flux produced in the galaxy. 
This quantity can be a probe of
the distribution and kinematics of the ISM.
For example, the outflowing ISM would make $f^{{\rm Ly}\alpha}_{\rm esc}$ 
larger since the number of resonant scattering and thus the chance of
absorption by dust are reduced \citep[e.g.,][]{kunth1998,atek2008}.
A clumpy distribution of the ISM would also 
make $f^{{\rm Ly}\alpha}_{\rm esc}$ larger 
\citep[e.g.,][]{hansen_oh2006,finkelstein2008}. 
In the clumpy ISM, H{\small II} regions where both Ly$\alpha$ and 
nebular lines originate are surrounded by clumpy gas clouds with 
well-mixed H{\small I} gas and dust 
(see, e.g., Fig. 1 of \citealt{neufeld1990}). 
We discuss the extent of clumpiness for our objects 
in \S \ref{subsec:clumpy_cloud}.

Assuming the Case B recombination where the intrinsic
Ly$\alpha$/H$\alpha$ ratio is 8.7 \citep{brocklehurst1971},
we estimate
the Ly$\alpha$ escape fraction as: 
\begin{equation}
f^{{\rm Ly}\alpha}_{\rm esc} \equiv \frac{L_{\rm obs}({\rm Ly}\alpha)} 
  { L_{\rm int}({\rm Ly}\alpha)}
 = \frac{L_{\rm obs}({\rm Ly}\alpha)} { 8.7L_{\rm int}({\rm H}\alpha)},
\end{equation}
where subscripts 'int' and 'obs' refer to intrinsic and
observed quantities, respectively,
and $L_{\rm int}({\rm H}\alpha)$ is obtained by correcting
the observed H$\alpha$ luminosity for dust extinction.
We derive the $L_{\rm obs}({\rm Ly}\alpha)$ of each object 
from its narrow and broad-band photometry on the assumption of
a flat continuum, 
taking accout of the exact position of the Ly$\alpha$ emission
in the NB387 response function  
and applying Madau (1995)'s prescription 
to correct for the IGM attenuation.
Note that these
$f^{{\rm Ly}\alpha}_{\rm esc}$ values are not measurements but
estimates, because we assume the Case B recombination.

We find
$f^{{\rm Ly}\alpha}_{\rm esc}$ $= 0.14\pm 0.03$ (CDFS-3865),
$0.29\pm 0.20$ (CDFS-6482),
$0.57\pm 0.13$ (COSMOS-13636),
and $0.17\pm0.03$ (COSMOS-30679).
These values are much higher than the average value
of $z\sim 2$ star-forming galaxies, $\sim5\%$ \citep{hayes2010}, 
but similar to the median value of 89 LAEs at $2< z < 4$, $\sim29\%$, 
obtained by \citet{blanc2011}.
It is interesting because our LAEs have relatively large
$E(B-V)$ values (Table \ref{tab:prop_sed}).
This would suggest that some mechanisms allow
Ly$\alpha$ photons to escape from the moderately
dusty ISM \citep{atek2008}.

\subsection{Size} \label{subsec:size}
Assuming that stars dominate the total mass of the luminous
(i.e., H$\alpha$ emitting) part of galaxies,
we infer the size of our objects from the virial theorem as:
\begin{equation}
r = G\frac{M_{*}}{\sigma_{\rm v}({\rm H}\alpha) ^{2}} 
\sim 4.3 \times
\frac{M_{*}/10^{10} M_{\odot}}
{(\sigma_{\rm v}({\rm H}\alpha)/100\ {\rm km\ s}^{-1}) ^{2}}\  
({\rm kpc}),
\end{equation}
where $M_{*}$ is the stellar mass and
$\sigma_{\rm v}({\rm H}\alpha)$ is the velocity dispersion 
measured from the H$\alpha$ line, 
corrected for the instrumental velocity dispersion 
($\sigma_{\rm inst.} = 86 \ [91]$ km s$^{-1}$ for NIRSPEC [MMIRS]).
It is known that for local galaxies 
stellar masses dominate total masses within effective radii 
(see e.g., Fig 2.a in \citealt{van_dokkum2009})
and that for high-$z$ galaxies 
total masses can be larger than stellar masses 
up to a factor of several (\citealt{erb2006b,van_dokkum2009}). 
Thus, our size estimates appear to be reasonable approximations 
of the true values within a factor of two or so.

\begin{figure}
\begin{center}
\includegraphics[scale=0.45]{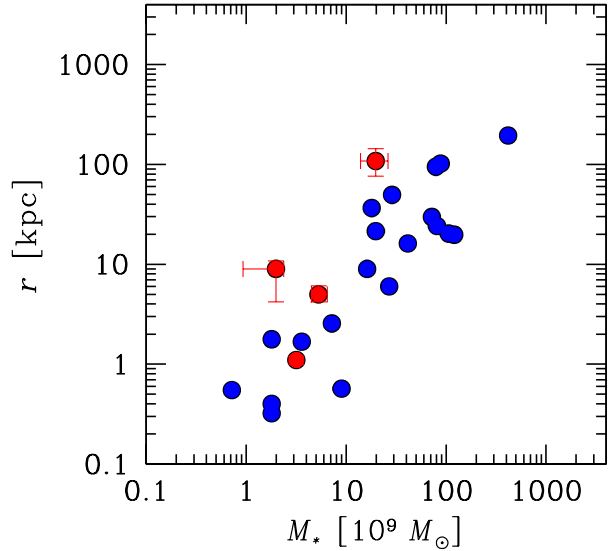}
\end{center}
\caption[]
{
Size plotted against stellar mass.
The red circles denote our LAEs, while
the blue circles represent 20 LBGs 
which have reliable measurements of stellar mass
and H$\alpha$ velocity dispersion 
\citep{erb2006b,steidel2010}.}

\label{fig:mass_size}
\end{figure}

We obtain $r=1.1^{+0.1}_{-0.1}$ kpc (CDFS-3865), 
$5.0^{+1.1}_{-0.8}$ kpc (CDFS-6482),
$9.0^{+1.8}_{-4.8}$ kpc (COSMOS-13636), 
and $107.5^{+35.6}_{-31.6}$ kpc (COSMOS-30679).
Similarly, we derive sizes for 20 LBGs of \citet{erb2006b}
which have measurements of stellar mass
and H$\alpha$ velocity dispersion.
Here we multiply the stellar masses of Erb et al.'s objects by $1.8$, 
since they assume a Chabrier IMF \citep{chabrier2003}.
As found from Figure \ref{fig:mass_size}, 
the LAEs and LBGs roughly lie on a single size-stellar mass relation,
suggesting that typical LAEs are smaller than LBGs 
just because they are less massive.

The physical quantities presented in this section are summarized in 
Table \ref{tab:summary_prop}

\section{Discussion} \label{sec:discussion}
\subsection{$\Delta$ $v_{Ly\alpha}$ and the Gas Motions of LAEs} 
\label{subsec:outflow}
Due to the resonant nature of Ly$\alpha$,
the observed Ly$\alpha$ line of a galaxy has a complicated
profile depending on the kinematics and geometry of the ISM.
A Ly$\alpha$ source in a simple static gas cloud produces
a symmetric double-peaked profile centered at 1216 \AA\
due to significant resonant scattering at 
1216 \AA \ \citep{harrington1973,neufeld1990,dijkstra2006}.
If the bluer peak is heavily absorbed by the intervening IGM
along the line of sight, only the redder peak will be observed.
In the case of outflow, 
the Ly$\alpha$ emission line should show an asymmetric profile
similar to a P Cygni profile.
\cite{verhamme2006} have explained some of observed Ly$\alpha$
profiles with a strong red peak and a weak blue peak with models
of an expanding shell
that absorbs Ly$\alpha$ photons at around 1216 \AA,
although the surface brightness distribution may not be explained by
such wind shells (see, e.g., \citealt{barnes2009}).
In short, when an outflow exists, the observed Ly$\alpha$ line 
should show an asymmetric profile with a strong peak 
redshifted with respect to the systemic velocity.

We measure the offset velocity of the Ly$\alpha$ peak,
$\Delta v_{\rm Ly\alpha}$, in \S \ref{subsec:velocity_offset}.
To quantify the asymmetry of the Ly$\alpha$ line of our objects,
we use the weighted skewness, $S_{w}$ \citep{kashikawa2006}.
Skewness is the third moment of the flux
distribution, and the weighted skewness
is defined as the product of the skewness and the line width.
A positive $S_{w}$ means that the Ly$\alpha$ profile has 
a red tail as in the case of outflow.
We obtain $S_{w} = 6.03\pm 0.51$ (CDFS-3865),
$5.45\pm 1.59$ (CDFS-6482),
$1.01\pm 0.76$ (COSMOS-13636), and
$5.31\pm 4.06$ (COSMOS-30679)
\footnote{
The $S_{w}$ of COSMOS-30679 is calculated after masking
the wavelength range affected by the cosmic ray.
}.

As found in \S \ref{subsec:velocity_offset},
three out of the four objects have a positive $\Delta v_{\rm Ly\alpha}$
beyond the $2\sigma$ uncertainty.
This result, combined with the positive measurements of $S_{w}$, 
leads us to a conclusion that the Ly$\alpha$ emission
of our objects is mostly originated from outflowing gas.
This conclusion is supported by the fact obtained in
\S \ref{subsec:metal_lines} that the LIS absorption lines 
of the composite spectrum are blue-shifted with respect to 
the systemic velocity.

Using simulations of Ly$\alpha$ radiative transfer
in outflowing galaxies,
\cite{verhamme2006,verhamme2008} suggest that not only LIS absorption
lines but also the Ly$\alpha$ line can be used to estimate
the outflow velocity.
Their simulations assume that a galaxy is surrounded by
a spherically symmetric shell-like outflowing gas cloud
in which {\sc Hi} gas and dust are well mixed.
They find that 
for relatively low {\sc Hi} column densities 
of $N_{\rm H} \lesssim 10^{20}$ cm$^{-2}$,
the peak of the Ly$\alpha$ profile emerges 
near the outflow velocity, 
giving $\Delta v_{\rm Ly\alpha} \sim v_{\rm out}$, 
while for high column densities of 
$N_{\rm H} \gtrsim 10^{20}$ cm$^{-2}$,
the peak is offset twice the outflow velocity, 
giving $\Delta v_{\rm Ly\alpha} \approx 2 v_{\rm out}$.
In any case, it is suggested that $\Delta v_{\rm Ly\alpha}$
positively correlates with $v_{\rm out}$.

On a reasonable assumption that $\Delta v_{\rm Ly\alpha}$ 
positively correlates with the speed of an outflow within a factor of 2, 
one can test the hypothesis that Ly$\alpha$ photons 
can escape more easily for a larger outflow velocity,
due to the reduced number of resonant scattering 
(e.g., \citealt{kunth1998}).
If this is the case, we should find a positive correlation
between $\Delta v_{\rm Ly\alpha}$ and EW(Ly$\alpha$).
However, Figure \ref{fig:dv_ew} indicates an opposite tendency. 
Furthermore, we find in \S \ref{subsec:metal_lines} that 
LAEs and LBGs have similar average $\Delta v_{\rm abs}$, 
i.e., similar average outflow velocities.
These two findings would suggest that outflows are not 
the physical origin of large-${\rm EW}({\rm Ly}\alpha)$ objects.

In \S \ref{subsec:velocity_offset}, we find that
the average $\Delta v_{\rm Ly\alpha}$ of LAEs is 
$175\pm 35$ km s$^{-1}$, 
which is significantly smaller than that of LBGs 
($\simeq 400$ km s$^{-1}$) at similar redshifts.
This result is important not only for understanding
the physical origin of Ly$\alpha$ emission in galaxies, 
but also for probing cosmic reionization with LAEs.
If LAEs at $z>6$ have similarly small $\Delta v_{\rm Ly\alpha}$ values,
the amount of Ly$\alpha$ photons scattered by the IGM, 
as used to constrain the epoch of reionization, may be in need of revision.
For example, \citet{santos2004}
has examined the transmission through the IGM 
of Ly$\alpha$ photons emitted from a galaxy for two cases, 
$\Delta v_{\rm Ly\alpha}=0$ and $360$ km s$^{-1}$, 
the latter of which is comparable to the average 
$\Delta v_{\rm Ly\alpha}$ of $z \sim 2$ -- 3 LBGs.
Some recent reionization studies using LAEs assume 
the latter case 
to estimate the neutral hydrogen fraction of the IGM, $x_{\rm HI}$, 
at $z > 6$ 
(e.g., \citealt{kashikawa2006,ota2008,ouchi2010,kashikawa2011}).
However, if $z>6$ LAEs have $\Delta v_{\rm Ly\alpha}$ as small as 
$\simeq 175$ km s$^{-1}$,
these studies may be overestimating $x_{\rm HI}$.
Similarly, \citet{ono2012}, \citet{schenker2012}, 
and \citet{pentericci2011}
have derived $x_{\rm HI}$ as large as $\sim 40$ -- $60$\% 
from a significant drop in the fraction of large-EW(Ly$\alpha$) 
galaxies from $z\sim 6$ to $7$. 
If such a high value were correct, reionization would take place very late,
which cannot easily be reconciled with constraints from the Lyman
alpha forest opacity \citep{becker2007} or the large value of
the Thomson optical depth, $\tau=0.09$,
obtained by WMAP observations \citep{dunkley2009,komatsu2011}.
Future Ly$\alpha$ emission models for 
reionization studies would need to use 
our result of a small average $\Delta v_{\rm Ly\alpha}$ value
if $\Delta v_{\rm Ly\alpha}$ does not significantly evolve  
over cosmic time.

\subsection{Correlations Between Outflow Velocity and
Physical Properties}\label{subsec:correlations_outflow}
\subsubsection{Outflow Velocity and SFR}\label{subsec:outflow_sfr}

We first examine the correlation between $\Delta v_{\rm Ly\alpha}$, 
which represents the outflow velocity within a factor of two as discussed 
in \S \ref{subsec:outflow},
and the SFR for LAEs and LBGs.
For the present-day universe, \cite{martin2005}
has found in a sample of ULIRGs, LIRGs, and starburst
dwarfs that those with a larger outflow velocity tend
to have a higher SFR, roughly following a power-law of 
$v_{\rm out}\propto SFR^{0.35}$. 
Note that this sample spans four orders of magnitude in the SFR. 
This correlation implies that
galaxies with a high SFR tend to have 
more massive stars and SNe which drive outflows.
As pointed out by some authors
(e.g., \citealt{martin2005,steidel2010}),
the correlation flattens and is difficult to see for objects 
with intermediate SFRs of $10$--$100\ M_\odot$ yr$^{-1}$.

From the left panel of Figure \ref{fig:dv_sfr}, 
there may exist a positive correlation between
$\Delta v_{\rm Ly\alpha}$ and SFR for LAEs, 
which is roughly consistent with 
$\Delta v_{\rm Ly\alpha} \propto SFR$, 
although the statistics is not very good.
For LBGs, which are taken
from the literature \citep{erb2006b,steidel2010}, 
a much flatter correlation is seen.
Thus, it is found that LAEs and LBGs do not follow the same power law,
and that LAEs have 
comparable to or systematically smaller outflow velocities
at a given SFR, even when we take account of 
the possibility that $\Delta v_{\rm Ly\alpha}$ may overestimate 
$v_{\rm out}$ up to a factor two.
Theoretically, the slope of the correlation depends on
the dominant physical outflow process \citep{kornei2012}.
A linear correlation, $v_{\rm out} \propto SFR$, is expected
for radiation pressure dominant outflows \citep{sharma2011},
while shallower slopes are indicative of
ram pressure dominant outflows \citep{heckman2000}.
It appears that outflows in LAEs
are at least consistent with the radiation pressure case.

\subsubsection{Outflow Velocity and $\sigma_{\rm v}({\rm H}\alpha)$}
\label{subsec:outflow_sigma}

We then examine how $\Delta v_{\rm Ly\alpha}$ correlates with
$\sigma_{\rm v}({\rm H}\alpha)$, 
which is related to the gravitational potential of galaxies.
\cite{martin2005} has found that the terminal outflow velocity
always approaches the galactic escape velocity,
which is consistent with a theoretical prediction
(e.g., \citealt{murray2005}).
The middle panel of Figure \ref{fig:dv_sfr} presents
$\Delta v_{\rm Ly\alpha}$ as a function of $\sigma_{\rm v}({\rm H}\alpha)$
for our LAEs,
together with those of $\sim 20$ LBGs \citep{erb2006b,steidel2010}.
It is not clear whether there exists a correlation
with the given poor statistics and large errors. 
However, LAEs are found to have comparable to or smaller outflow velocities
than LBGs at a given $\sigma_{\rm v}({\rm H}\alpha)$.

\subsubsection{Outflow Velocity and SFR Surface Density}
\label{subsec:outflow_sfrdensity}

\begin{figure*}
\begin{center}
\includegraphics[width=18cm]{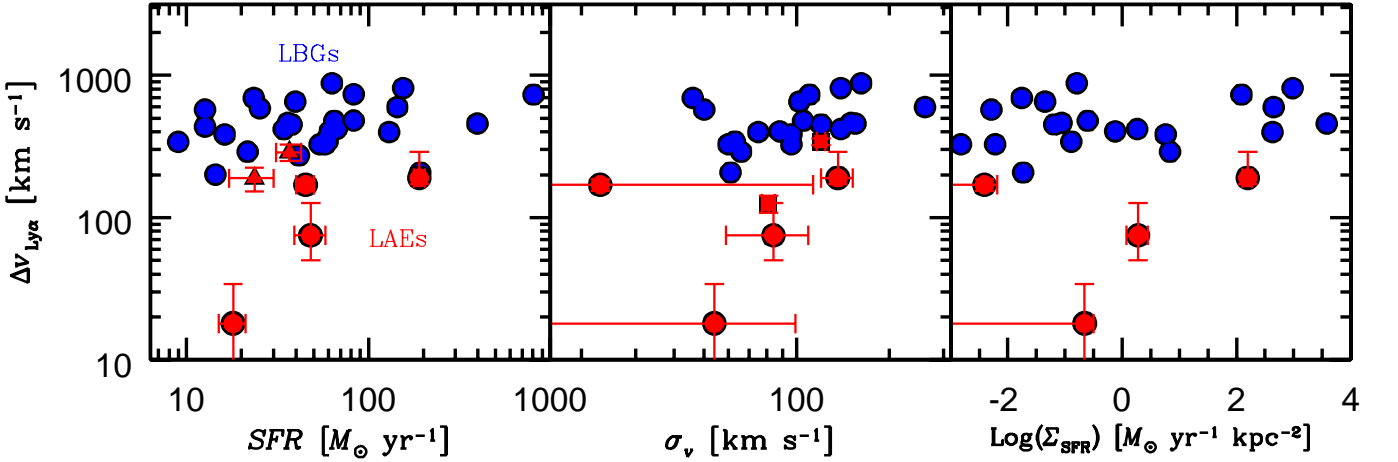}
\end{center}
\caption[]
{
$\Delta v_{\rm Ly\alpha}$ as a function of SFR (left panel),
$\sigma_{\rm v}({\rm H}\alpha)$ (middle), 
and $\Sigma_{\rm SFR}$ (right).
For all panels, the red and blue circles indicate, respectively, 
our LAEs and LBGs taken from the literature. 
The red triangles in the left panel denote
LAEs at $z\sim2.3$ \citep{finkelstein2011b}, 
and the red squares in the middle panel indicate
LAEs at $z\sim3$ \citep{mclinden2011}.
}
\label{fig:dv_sfr}
\end{figure*}

Finally, we examine the correlation
with the SFR surface density, $\Sigma_{\rm SFR}$, 
which represents the intensity of star forming activity.
The right panel of Figure \ref{fig:dv_sfr} plots
$\Delta v_{\rm Ly\alpha}$ against $\Sigma_{\rm SFR}$ 
for our LAEs and LBGs taken from the literature.
Here we estimate $\Sigma_{\rm SFR}$ by dividing the SFR by
$\pi r^2$, where $r$ is the size derived in \S \ref{subsec:size}.
The $\Sigma_{\rm SFR}$ values of our objects are
$157^{+31}_{-31}\ M_{\odot}$ yr$^{-1}$ kpc$^{-2}$ (CDFS-3865),
$1.9^{+0.9}_{-0.7}\ M_{\odot}$ yr$^{-1}$ kpc$^{-2}$ (CDFS-6482), 
$0.22^{+0.096}_{-0.22}\ M_{\odot}$ yr$^{-1}$ kpc$^{-2}$ (COSMOS-13636), 
and 
$3.89^{+2.62}_{-2.33} \times 10^{-3}\ M_{\odot}$ yr$^{-1}$ kpc$^{-2}$ 
(COSMOS-30679).
\footnote{
Although many studies have adopted Petrosian radius
for the size of a galaxy, it is known that Petrosian radius is
likely to overestimate the area of the star forming region 
of galaxies
especially for high-$z$ objects with a clumpy star forming region 
\citep{kornei2012}.
}
Again, LAEs are found to have comparable to or smaller 
outflow velocities than LBGs at a given $\Sigma_{\rm SFR}$.

To summarize, we find that LAEs tend to have comparable to or 
smaller outflow velocities than LBGs 
and that LAEs and LBGs have different slopes of 
the $\Delta v_{\rm Ly\alpha}$-SFR relation.
These findings imply that the physical origin of LAEs' outflows
appears to be different from that of LBGs'.

\subsection{Why LAEs Have Strong Ly$\alpha$ Emission?}
\label{subsec:why_strong}

In this subsection, we discuss
the physical origin of strong Ly$\alpha$ emission from LAEs.
There are various mechanisms capable of enhancing
Ly$\alpha$ emission: 
shock heating by an inflow or outflow, very weak dust extinction, 
a peculiar geometry of dust/gas clouds, 
a low neutral hydrogen column density, 
and a high outflow velocity. 
The high outflow velocity hypothesis is 
found to be unlikely in \S \ref{subsec:outflow} 
(see also the left panels of Figure \ref{fig:ew_f_all}).
Here we examine the weak-dust extinction hypothesis, 
clumpy-cloud hypothesis, 
and low-$N_{\rm H}$ hypothesis.

\subsubsection{Weak Dust Extinction}\label{subsec:screen_dust}

It is possible that Ly$\alpha$ photons survive
through resonant scattering in gas clouds with little dust.
In this case, one should find significantly small $E(B-V)$
in LAEs and a negative correlation between $E(B-V)$
and EW(Ly$\alpha$).
The middle panels of Figure \ref{fig:ew_f_all} plot 
EW(Ly$\alpha$) and $f_{\rm esc}^{{\rm Ly}\alpha}$ as functions of $E(B-V)$.
There is no significant correlation in either panel.
\cite{hayes2010} and \cite{kornei2010} have reported an anti-correlation
between $f_{\rm esc}^{{\rm Ly}\alpha}$ and $E(B-V)$ in H$\alpha$ emitters
at $z\simeq 2.2$ and LBGs at $z\sim3$, respectively.
However, in our $f_{\rm esc}^{{\rm Ly}\alpha}$ vs $E(B-V)$ plot,
there are objects with relatively high
$f_{\rm esc}^{{\rm Ly}\alpha}$ of $0.3$--$0.6$ even in 
a moderately high extinction of $E(B-V)=0.2$--$0.3$.
Therefore, it is not clear if weak dust extinction solely can explain
the strong Ly$\alpha$ emission.

\subsubsection{Clumpy Clouds}\label{subsec:clumpy_cloud}

The gas distribution in LAEs may not be smooth
and spherically symmetric, but clumpy.
In a clumpy geometry, dust grains are shielded by H{\small I} gas,
and Ly$\alpha$ photons are resonantly scattered on the surfaces 
of clouds without being absorbed by dust
\citep{neufeld1991,hansen_oh2006}.
Because continuum photons are absorbed through dusty 
gas clouds, the ratio of Ly$\alpha$ to UV continuum fluxes,
or EW(Ly$\alpha$), is enhanced.
For a further discussion of the clumpy-cloud hypothesis,
we calculate the clumpiness parameter,
$q$, introduced by \cite{finkelstein2008}:
\begin{equation}
q = \tau({\rm Ly}\alpha)/\tau_{1216},
\end{equation}
where $\tau({\rm Ly}\alpha)$ and $\tau_{1216}$ are defined as
$e^{-\tau({\rm Ly}\alpha)} = L_{\rm obs}({\rm Ly}\alpha)/L_{\rm int}({\rm Ly}\alpha)$
and $e^{-\tau_{1216}} = 10^{-0.4k_{1216}E(B-V)}$
with the extinction coefficient at $\lambda = 1216$ \AA,
$k_{1216}=11.98$ \citep{calzetti2000}.
This parameter is used to diagnose the geometry of the ISM
\citep[e.g.,][]{finkelstein2008,kornei2010,nakajima2012a}.
If the geometry of the ISM is clumpy,
$q$ is smaller than unity.
In the case of $q>1$, Ly$\alpha$ photons are more preferentially absorbed by dust
through the relatively homogeneous ISM.

The $q$ values of our objects are calculated to be
$q=0.96\pm0.11$ (CDFS-3865), $0.61\pm0.34$ (CDFS-6482),
$0.19^{+0.07}_{-0.09}$ (COSMSO-13636),
and $0.30^{+0.03}_{-0.03}$ (COSMOS-30679);
three out of the four clearly have $q<1$ beyond the uncertainties.
These results are consistent with that of
\cite{nakajima2012a} who obtained $q=0.7^{+0.1}_{-0.1}$
for $z \simeq 2.2$ stacked LAEs. On the other hand,
some studies have found more LAEs with $q>1$
at $z\sim2$ \citep[e.g.,][]{nilsson2009,hayes2010}.

\begin{figure*}
\begin{center}
\includegraphics[scale=0.8]{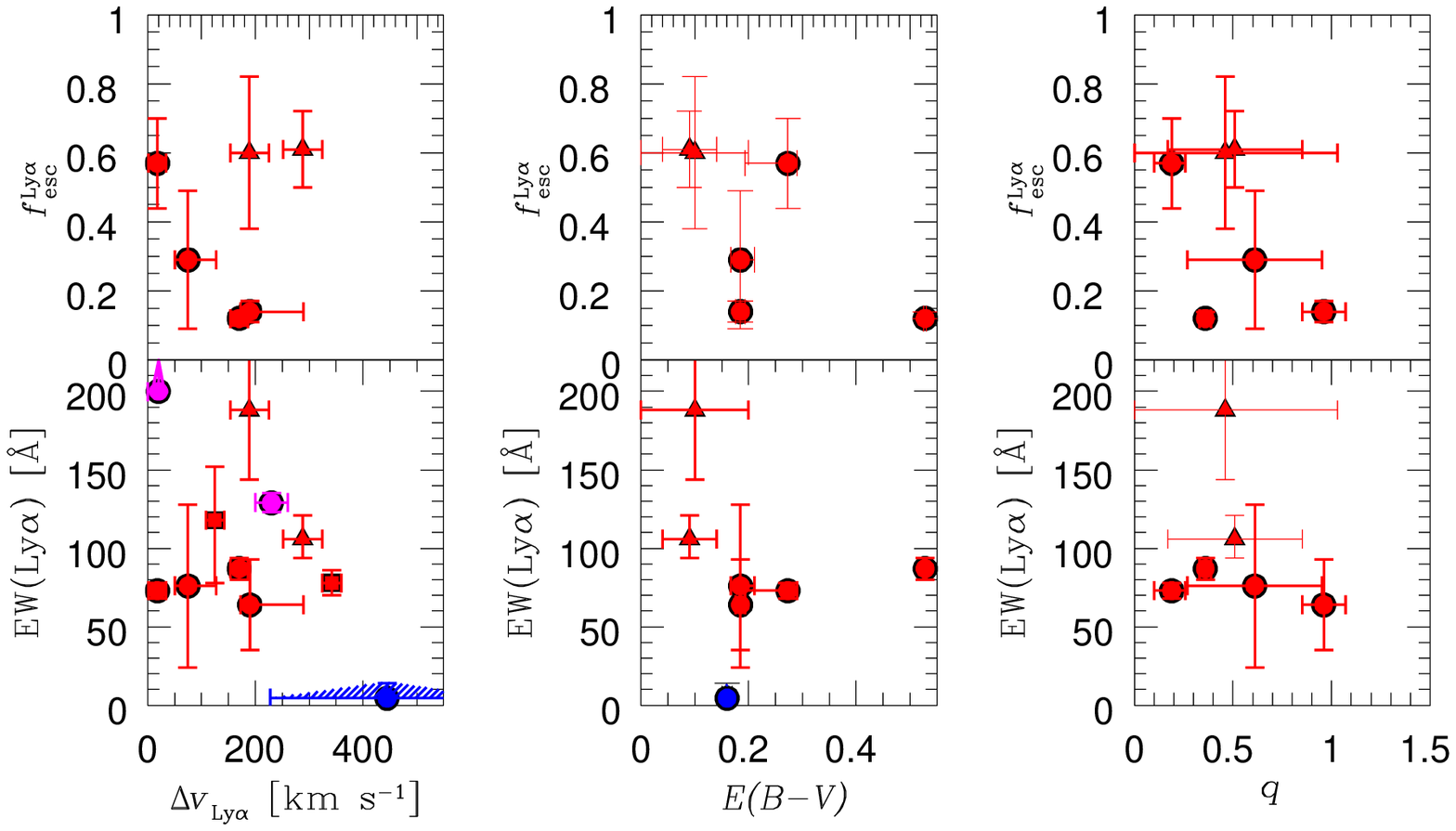}
\end{center}
\caption[]
{
$f_{\rm esc}^{{\rm Ly}\alpha}$ and EW(Ly$\alpha$) plotted against 
$\Delta v_{\rm Ly\alpha}$, $E(B-V)$, and $q$.
The red circles denote our LAEs. 
The red squares and triangles are the LAEs
given by \cite{mclinden2011} and \cite{finkelstein2011b}, respectively.
The magenta circles and blue circles represent, respectively, 
LABs \citep{yang2011} and LBGs \citep{erb2006b,reddy2008,steidel2010}.
}
\label{fig:ew_f_all}
\end{figure*}

The right panels of Figure \ref{fig:ew_f_all} show
$f_{\rm esc}^{{\rm Ly}\alpha}$ and EW(Ly$\alpha$) as functions of $q$
for the LAE sample.
There may exist a negative correlation between $q$ and
$f_{\rm esc}^{{\rm Ly}\alpha}$, although not very clear due
to the poor statistics.

\subsubsection{Low $N_{\rm H}$}\label{subsec:small_N}

Finally, we examine the possibility that the $N_{\rm H}$ 
of LAEs is relatively low.
To constrain $N_{\rm H}$ for LAEs and LBGs,
we use the dependence of $\Delta v_{\rm Ly\alpha}$ on 
$N_{\rm H}$ found by \cite{verhamme2006,verhamme2008}.
As mentioned in \S \ref{subsec:outflow},
they find 
$\Delta v_{\rm Ly\alpha} \approx 2 v_{\rm out}$ 
for $N_{\rm H} \gtrsim 10^{20}$ cm$^{-2}$ 
while $\Delta v_{\rm Ly\alpha} \sim v_{\rm out}$ 
for $N_{\rm H} \lesssim 10^{20}$ cm$^{-2}$;
$\Delta v_{\rm Ly\alpha}$ is smaller for lower $N_{\rm H}$ 
due to a smaller amount of wavelength shift 
by resonant scattering.
Since $|\Delta v_{\rm abs}|$ is equal to $v_{\rm out}$  
in an expanding shell, 
one can thus use $\Delta v_{\rm Ly\alpha}$ and $\Delta v_{\rm abs}$ 
to distinguish between high and low {\sc Hi} column densities.

The LAEs we study have average velocities of 
$\Delta v_{\rm Ly\alpha} = 175\pm35$ km s$^{-1}$ 
and $\Delta v_{\rm abs} = -179\pm73$ km s$^{-1}$,
i.e., $\Delta v_{\rm Ly\alpha} \approx |\Delta v_{\rm abs}|$, 
while LBGs have
$\Delta v_{\rm Ly\alpha} \approx 2$--$3 \times |\Delta v_{\rm abs}|$
(\citealt{steidel2010,rakic2011}). 
This suggests that LAEs have on average lower $N_{\rm H}$
than LBGs.

For a lower $N_{\rm H}$, 
Ly$\alpha$ photons have a reduced chance of 
absorption by dust before escaping the galaxy
because of a smaller number of resonant scattering.
Thus, we infer that LAEs have strong Ly$\alpha$ emission 
because of low {\sc Hi} column densities.
Although our data tell nothing about why LAEs have 
low column densities, possible reasons would include 
a high gas temperature caused by a shock heating 
(\citealt{nakajima2012b}), 
which reduces the fraction of neutral hydrogen, 
and a clumpy gas geometry (\citealt{bond2010}).

\begin{deluxetable*}{ccccccccccc}
\tablecolumns{11}
\tablewidth{0pt}
\tablecaption{Summary of the physical quantities \label{tab:summary_prop}}
\tablehead{
\colhead{Object} & \colhead{$E(B-V)_{*}$} & \colhead{$M_{*}$} & \colhead{EW(Ly$\alpha$)}
& \colhead{$\Delta v_{\rm Ly\alpha}$} & \colhead{$ L({\rm H}\alpha)$} & \colhead{$\sigma_{\rm v} ({\rm H}\alpha)$}
& \colhead{$SFR({\rm H}\alpha)$} & \colhead{$f^{{\rm Ly}\alpha}_{\rm esc}$} & \colhead{$r$} & \colhead{$q$}
\\
\colhead{} &\colhead{} & \colhead{$10^{9} M_\odot$} & \colhead{\AA} & \colhead{km s$^{-1}$}
& \colhead{10$^{41}$ erg s$^{-1}$} & \colhead{km s$^{-1}$}
& \colhead{$M_\odot$ yr$^{-1}$} & \colhead{} & \colhead{kpc} & \colhead{}
\\
\colhead{} &\colhead{(1)} & \colhead{(2)} & \colhead{(3)} & \colhead{(4)}
& \colhead{(5)} & \colhead{(6)}
& \colhead{(7)} & \colhead{(8)} & \colhead{(9)} & \colhead{(10)}
}\\
\startdata
\hline \\
CDFS-3865
&  $0.185^{+0.009}_{-0.009}$ &  $3.18^{+0.21}_{-0.13}$ & $64^{+29}_{-29}$
& $190^{+99}_{-18}$ &  $136.5^{+8.5}_{-8.5}$ &$110.8$  & $190^{+13}_{-13}$
&  $0.14^{+0.03}_{-0.03}$ & $1.1^{+0.1}_{-0.1}$ & $0.96^{+0.11}_{-0.11}$
\\
CDFS-6482
&  $0.185^{+0.026}_{-0.018}$ & $5.30^{+1.18}_{-0.80}$ & $76^{+52}_{-52}$
& $75^{+52}_{-25}$ &  $34.5^{+6.2}_{-6.2}$  &  $67.5$ &  $48^{+10}_{-9}$
& $0.29^{+0.20}_{-0.20}$ & $5.0^{+1.1}_{-0.8}$ & $0.61^{+0.34}_{-0.34}$
\\
COSMOS-13636
&  $0.273^{+0.018}_{-0.079}$ &  $1.99^{+0.39}_{-1.06}$ & $73^{+5}_{-5}$
& $18^{+16}_{-16}$ &  $9.5^{+1.3}_{-1.3}$ &  $30.7$ &  $18^{+3}_{-3}$
&  $0.57^{+0.13}_{-0.13}$ & $9.0^{+1.8}_{-4.8}$ & $0.19^{+0.07}_{-0.09}$
\\
COSMOS-30679
&  $0.528^{+0.0026}_{-0.0026}$ & $19.75^{+6.53}_{-5.80}$ & $87^{+7}_{-7}$ 
& $170^{+16}_{-16}$ &  $11.4^{+1.0}_{-1.0}$ &  $28.1$ &  $45^{+5}_{-5}$
& $0.17^{+0.03}_{-0.03}$& $107.5^{+35.6}_{-31.6}$ & $0.30^{+0.03}_{-0.03}$
\enddata

\tablenotetext{}{Notes.
(1) Dust extinction estimated from SED fitting;
(2) Stellar mass estimated from SED fitting;
(3) Rest-frame Ly$\alpha$ EW;
(4) Velocity offset of the Ly$\alpha$ line; 
(5) H$\alpha$ luminosity;
(6) Velocity dispersion of H$\alpha$ line;
(7) Star formation rate from H$\alpha$ luminosity corrected for dust extinction;
(8) Ly$\alpha$ escape fraction estimated in \S \ref{subsec:escape_fraction};
(9) Size of the galaxy derived in \S \ref{subsec:size} under the
assumption that stars dominate the total mass in the luminous part of the galaxy;
(10) Clumpiness parameter (\S \ref{subsec:clumpy_cloud}).
}

\end{deluxetable*}

\vspace{40pt}

\section{CONCLUSIONS} \label{sec:conclusions}

We have presented the results of Magellan/MMIRS and Keck/NIRSPEC
spectroscopy for five LAEs at $z \simeq 2.2$
for which high-resolution FUV spectra from Magellan/MagE 
are available.
These objects are taken from the $z \simeq 2.2$ LAE samples 
constructed by K. Nakajima et al. 
(in preparation: see \citealt{nakajima2012a} for the selection).
The redshift of $z \simeq 2.2$ is unique since we can observe 
from the ground all of Ly$\alpha$, LIS absorption lines, 
and optical nebular emission lines including H$\alpha$.

We have successfully detected H$\alpha$ emission from all five objects,
and {\sc [Oii]} $\lambda\lambda3726,3729$, H$\beta$, and/or 
{\sc [Oiii]} $\lambda\lambda4959,5007$ for some LAEs
on the individual basis.
In addition to that, we have detected 
LIS absorption lines in a stacked FUV spectrum.
After removing an AGN-contaminated object,
we have measured the velocity offsets of the Ly$\alpha$ line 
($\Delta v_{\rm Ly\alpha}$) 
and of LIS absorption lines ($\Delta v_{\rm abs}$)
from the systemic redshift determined by nebular emission lines,
to discuss the gas motions of LAEs.
The major results of our study are summarized below.

\begin{itemize}

\item
We have obtained 
$\Delta v_{\rm Ly\alpha}$ = $190^{+99}_{-18}$ km s$^{-1}$ (CDFS-3865),
$75^{+52}_{-25}$ km s$^{-1}$ (CDFS-6482),
$18\pm16$ km s$^{-1}$ (COSMOS-13636),
and $170\pm16$ km s$^{-1}$ (COSMOS-30679); 
three out of the four have a positive $\Delta v_{\rm Ly\alpha}$
beyond the $2\sigma$ uncertainty.
Combining with the result that all four have a positive 
weighted skewness, 
we have conclude that the Ly$\alpha$ emission of our objects 
is mostly originated from outflowing gas.
This conclusion is supported by the finding that 
the LIS absorption lines in the stacked FUV spectrum are 
blue-shifted with $\Delta v_{\rm abs} = -179 \pm73$ km s$^{-1}$.

\item
For a sample of eight $z \sim 2$--$3$ LAEs without AGN
from our study and the literature, 
we have obtained $\Delta v_{\rm Ly\alpha} = 175\pm35$ km s$^{-1}$, 
which is significantly smaller than that of LBGs, 
$\Delta v_{\rm Ly\alpha} \simeq 400$ km s$^{-1}$.
If LAEs at $z>6$ have similarly small $\Delta v_{\rm Ly\alpha}$ values,
some reionization studies based on LAEs 
assuming $\Delta v_{\rm Ly\alpha}$ as large as that of LBGs 
may be overestimating the neutral fraction of the IGM.

\item
We have found 
an anti-correlation between EW(Ly$\alpha$) and $\Delta v_{\rm Ly\alpha}$
in a compilation of LAE, LAB, and LBG samples, 
i.e., EW(Ly$\alpha$) decreases with $\Delta v_{\rm Ly\alpha}$.
On a reasonable assumption that $\Delta v_{\rm Ly\alpha}$ 
positively correlates with the outflow velocity within 
a factor of two, 
this anti-correlation indicates that 
high outflow velocities are not the physical origin 
of the strong Ly$\alpha$ emission of LAEs.

\item
We have found that LAEs have comparable to or smaller outflow
velocities than LBGs at a given $SFR$,
$\sigma({\rm H}\alpha)$, and $\Sigma_{\rm SFR}$  
(when the systematic error of $\lesssim 2$ in $\Delta v_{\rm Ly\alpha}$ 
as a measure of $v_{\rm out}$ is considered).
We have also found that the slope of the $\Delta v_{\rm Ly\alpha}$-SFR
relation is different between LAEs and LBGs.
Thus, the physical origin of LAEs' outflows appears to be different
from that of LBGs'.
It appears that LAEs' outflows 
are at least consistent with the radiation pressure case.

\item
To identify the physical origin of large-EW(Ly$\alpha$) galaxies, 
we have tested three hypotheses which may facilitate the escape of Ly$\alpha$ 
photons: weak dust extinction, clumpy-cloud geometry, 
and low $N_{\rm H}$.
Since we have found no significant correlation between EW(Ly$\alpha$)
and $E(B-V)$, it is not clear if weak dust extinction 
leads to a high escape fraction of Ly$\alpha$ photons.
Although there may exist an anti-correlation between 
$f_{\rm esc}^{{\rm Ly}\alpha}$ and the clumpiness parameter, 
more data and theoretical work are needed 
to draw a firm conclusion.

\item
We have found that 
LAEs have $\Delta v_{\rm Ly\alpha} \approx |\Delta v_{\rm abs}|$, 
in contrast with LBGs which have 
$\Delta v_{\rm Ly\alpha} \approx 2$--$3 \times |\Delta v_{\rm abs}|$.  
When combined with the simulations of Ly$\alpha$ radiative transfer 
in a galaxy with an expanding shell of the ISM 
by \citealt{verhamme2006,verhamme2008}, 
this suggests that the typical $N_{\rm H}$ of LAEs would be lower than 
that of LBGs, giving a smaller number of resonant scattering. 
Such low $N_{\rm H}$ may 
cause the observed strong Ly$\alpha$ emission of LAEs.

\end{itemize}

\section*{Acknowledgements}
We thank an anonymous referee for valuable comments that
have greatly improved the paper.
We are grateful to Kentaro Motohara and Masakazu Kobayashi for their helpful comments.
We acknowledge Brian McLeod and Paul Martini who
gave us helpful advice on our MMIRS observations and data reduction.
We are deeply grateful to Magellan and Keck Telescope staff 
for supporting our MMIRS and NIRSPEC observations. 
We also thank Yujin Yang for providing photometry data of two LABs, 
Steven Finkelstein for providing the revised results of his work, 
Daniel Stark, Ann Zabludoff, Masayuki Umemura, Hannes Jensen, and Matthew Schenker 
for giving us various useful comments.
This work was supported by KAKENHI (23244025) Grant-in-Aid
for Scientific Research(A)
through Japan Society for the Promotion of Science (JSPS),
and NSF grant 1108815 awarded by National Science Foundation.



\end{document}